\documentclass[prb,showpacs,superscriptaddress,amssymb,twocolumn,amsfonts,eqsecnum]{revtex4-1}
\usepackage{amsmath}
\usepackage{bm}
\usepackage{latexsym}
\usepackage{graphics}
\usepackage{dsfont}
\usepackage{amssymb}
\usepackage{tabularx}

\def\beq{\begin{equation}}
\def\eeq{\end{equation}}
\def\beqn{\begin{eqnarray}}
\def\eeqn{\end{eqnarray}}

\begin{document}
\title{Geometry  of Fractional Quantum Hall Fluids}

\author{Gil Young Cho, Yizhi You, and Eduardo Fradkin}
\affiliation{Department of Physics and Institute for Condensed Matter Theory, University of Illinois, 1110 W. Green St., Urbana IL 61801-3080, U.S.A.}

\date{\today}

\begin{abstract}
We use the field theory description of the fractional quantum Hall states to derive the universal response of these topological fluids to  shear deformations and curvature of their background geometry, i.e. the Hall viscosity, and  the Wen-Zee term. %the gravitational Chern-Simons term. 
To account for the coupling to the background geometry, we show that the concept of flux attachment needs to be modified and use it to derive the geometric responses from Chern-Simons theories. We show that the resulting composite particles minimally couple to the spin connection of the geometry. We derive a consistent theory of geometric responses from the Chern-Simons effective field theories and from parton constructions, and apply it to both abelian and non-abelian states.
\end{abstract}

%\pacs{71.35.-y, 71.10.Pm, 73.20.-r}

\maketitle

\section{Introduction}

The quantized Hall conductivity is the most fundamental transverse response of the incompressible fractional quantum Hall (FQH) states of two-dimensional electron fluids in external magnetic fields.\cite{Laughlin83,Zhang1992, Haldane1983,Jain1989,Lopez1991,Wen95,wenrev1,Wen2004book,fradkin-1991} The charge current flows perpendicular to the direction of an external in-plane electric field, and   the transport is dissipationless. The Hall conductivity does not depend on the microscopic details of the system, but only on the topological properties of the states. The Hall conductivity is one of the key topological properties  characterizing the quantum Hall fluids. However, the Hall conductivity  does not fully characterize these topological  fluids.\cite{wenrev1,Wen2012,Wen95}

A full characterization of abelian FQH states as topological fluids \cite{Wen95,Wen2012} includes the (fractional) charge and statistics of the quasiholes and the ground state degeneracy on closed surfaces, as well as various so-called fusion reels for the quasi-holes. These dimensionless universal properties of the FQH state  are determined by topological invariants of the topological fluid states. 
However, a full characterization of the FQH fluid also requires the  intrinsic orbital spin\cite{Wen95} $s$, and the associated Hall viscosity $\eta_{H}$. These quantities express   the way the fluid couples to geometric properties of the 2D surface on which it moves\cite{wenzee, Read2009, Read2011, Hoyos2012, Bradlyn2012, Haldane2009, Hughes2011, Hughes2013, Gromov2014, Gromov2014b, Nicolis2011, Son2013, Avron1995, Hansson2013}. They become manifest when a FQH state is put on a curved (and dynamical) surface. The Hall viscosity is the response of the Hall fluid to an external shear deformation of the background surface, under which the Hall fluid develops a momentum density perpendicular to  it. As a result, the net energy for the deformation vanishes, resulting in a non-dissipative viscosity.\cite{Avron1995}  It has been argued that the Hall viscosity $\eta_{H}$ depends only on the density of electron ${\bar \rho}$ and the orbital spin $s$ through the relation\cite{Read2009, Read2011,Hansson2013, Wen2012} 
$\eta_{H}= s {\bar \rho}/2$.

The coupling of the  Hall fluid  to the  curvature of the background surface is the origin of the shift vector associated to  FQH states on spheres 
 \cite{wenzee,Wen95}. At the level of the effective hydrodynamic theory this coupling is  represented in its effective action by  the Wen-Zee term (whose  coefficient  involves the orbital spin\cite{wenzee,Wen95, Read2009, Read2011, Haldane2009} $s$). It represents the universal coupling of the hydrodynamic gauge fields of the fluid to the spin connection of the geometry of the surface.  The Wen-Zee term was introduced to account for an additional Berry phase needed to represent the Hall  fluid  on a sphere, and also  predicts that local changes in the curvature of the surface should be accompanied by local accumulation of electric charge. 
Since the orbital spin $s$ and the geometric response are closely related to each other,  a calculation of the geometric response   amounts to a derivation of the orbital spin $s$ of the fluid (for a rotationally invariant  system \cite{Read2009, Read2011}).  

The topological properties of FQH fluids are encoded in  the effective hydrodynamic theory which has the form of Chern-Simons gauge theory.\cite{Wen1992,Wen95}
At a microscopic level, the FQH states are described either from the structure of model  wave functions
 \cite{Laughlin83,Halperin1984,Jain1989,Moore1991,Read1999}, Chern-Simons gauge  theories that implement  the concept of flux attachment,\cite{Zhang1989,Lopez1991} or by parton constructions.\cite{Wen1999}
In the past, the Hall viscosity and Wen-Zee term have been studied in various ways, ranging from the modular properties of FQH wave functions, using AdS/CFT holographic dual methods, to modeling hydrodynamic theories of FQH states (and in topological insulators.)\cite{wenzee,Read2009, Read2011, Hoyos2012, Bradlyn2012, Haldane2009, Hughes2011, Hughes2013, Gromov2014, Gromov2014b, Nicolis2011, Son2013, Avron1995, Hansson2013,saremi-2012,son-2013-holographic,chen-2012a,chen-2012b} However, so far there has been no consistent derivation of the geometrical properties of FQH fluids from their field theoretic descriptions. 

In this paper, we derive the Hall viscosity and the Wen-Zee term using the description of Chern-Simons gauge theories, which embody the concept of flux attachment, and also with the projective parton approach.\cite{Wen1999, Barkeshli2010} To derive the geometric response from the Chern-Simons gauge theories, we first show that the conventional approach to  flux attachment must  be modified even for for a system of non-relativistic particles moving in a curved space. 
We show that the resulting composite particles are minimally coupled to the spin connection of the geometry (even though the microscopic particles are  scalars.)  The strength of this coupling is identified with the topological spin induced by the flux attached to the electron. We show that the coupling to the spin connection is essential to reproduce the geometric responses of FQH states.  We  also derive the effective field theory using the parton construction,\cite{Wen1999, Barkeshli2010}  including the geometric responses for general FQH states including non-abelian states. We get a consistent understanding of two-dimensional abelian topological orders\cite{Wen2012} from the effective field theoretic approaches as well as geometric responses of more general class of the topological states including non-abelian FQH states. 

We then further show that the  {\it straightforward} application of the composite particle theories and projective parton constructions {\it fail} to reproduce the correct CS action for the spin connection, i.e. the gravitational CS action, whose coefficient should be equal to the central charge of the Virasoro algebra of the edge states of the fluid on a disk geometry. The gravitational CS term\cite{Deser-1982,Witten1988,Witten1989,witten-2007} reflects the gravitational anomaly\cite{Alvarez-Gaume-1984} of the energy-momentum tensor in topological fluids.\cite{Ryu-2012,Stone-2012,Luttinger-1964} We find that the gravitational CS term resulting from the field theoretic descriptions predicts the central charge of the mean field theory used in the descriptions, instead of the correct ground state. For example, the correct central charge of the Laughlin state is $c=1$ because of a single chiral edge state of the state. However, the composite boson theory predicts the central charge of the Laughlin state to be zero because the composite boson theory describes the state as the  (approximately) time-reversal symmetric superfluid state. We will show that the composite fermion theory and the projective parton construction suffers the same problem on predicting the correct central charge. Thus, predicting the gravitational CS term of FQH states using the field theoretic approaches remains an open problem.  

%In addition, the 
%The coupling to the spin connection leads to an extra topological term in the response: the (gravitational) CS action for the 
%spin connection\cite{Deser-1982,witten-1988b,witten-2007}
%. The significance of this term is that it 
%which yields the energy-momentum tensor of the fluid. 
%Its 

%, {\it e.g.,} ${\mathbb Z}_{k}$ parafermion state. 
%Our results yield a 
%(except for some  issues on the gravitational CS term.)
% in the latter case.)  

This paper is organized as follows. In Section \ref{sec:flux-attachment} we show that on a curved surface (a manifold), a consistent theory of flux attachment necessarily requires to take into account the topological spin. Here we derive the form of the resulting Chern-Simons gauge theory which now includes a coupling to the spin connection of the manifold. In Section \ref{sec:CF} we use this theory to the case of the composite fermion construction of the Laughlin and Jain states and to the multi-component abelian FQH states. Here we derive the effective hydrodynamic theories for each case and show that they now predict the correct value of the Hall viscosity and of the Wen-Zee term in each case. However we also find that in general the composite particle theories do not predict the correct value of the coefficient of the gravitational Chern-Simons term which should be consistent with the value of the central charge of the theory of the chiral edge states.  In Section \ref{sec:CB} we present the equivalent description for the theory of composite bosons and in Section \ref{sec:parton} we extend this formulation to the parton construction of abelian and non-abelian FQH states. Our conclusions are presented in Section \ref{sec:conclusions}.

\section{Flux attachment and Geometry} 
\label{sec:flux-attachment}

In the descriptions of FQH states, the Chern-Simons (CS) term plays an important role: it binds the flux to the charge and induces the statistical transmutation. Another equally important but less appreciated ingredient from the CS term is the topological spin.\cite{Hansson1988Spin,Grundberg1989, Polyakov1988,Tze1988Manifold,Semenoff1989,Dunne1989} Formally, the spin can be introduced 
 %to the theory 
 by defining a local frame attached to the worldline of the charge-flux composite particle.
 % which is 
 %done by formally regarding the worldline of a composite particle 
 %as a ``ribbon''
 % the ends of the ribbon 
 %corresponding to the positions of the charge and the flux bound to the charge. 
 %Then the ribbon will be twisted as the composite particle moves in spacetime. 
 The topological spin  counts %how fast 
 the winding of the charge %circulates 
 around the flux. 
 %This construction 
 %is widely applicable  
 %applies to all systems in $D=2$ space dimensions, even without  an external magnetic field.

The spin is related with the self-statistical angle of the composite particle. To be precise, we consider a CS gauge theory minimally coupled to the charge current $j^{\mu}$,  
\beq
{\mathcal L} = \frac{k}{4\pi} \varepsilon^{\mu\nu\lambda} a_{\mu} \partial_{\nu} a_{\lambda} - j^{\mu}a_{\mu}
\label{eq:CS}
\eeq
The content of the CS Lagrangian, Eq.\eqref{eq:CS},  is a charge-flux constraint (the Gauss law of this theory) and canonical commutation relations for the gauge fields \cite{Witten1989}.
The charge is bound with the flux and is turned into a composite particle with the change in the statistical angle $\theta_{\rm stat}= \frac{\pi}{k}$. Then the spin-statistics connection implies that the composite particle will carry the topological spin $S_{z} = \frac{\theta_{\rm stat}}{2\pi} = \frac{1}{2k}$. Thus the composite particle carries  spin polarized along the ${\bm z}$-direction.

On a surface with a non-flat metric  the topological spin of the composite particle couples to the (abelian) spin connection with a coupling strength dictated by the topological spin.  
To demonstrate this, we perform a parallel transport of a composite particle along the curve $C: s \rightarrow {\bm r} = (x_{1}(s), x_{2}(s), t(s)) \in \Sigma^{2} \times {\mathbb R}$ with its arc length $s$. We are interested in the adiabatic transport of the particle, {\it i.e.,} $|\frac{d {\bm x}}{ds}|^{2} << |\frac{dt}{ds}|^{2}$ along $C$. In the pure CS theory, the amplitude for the transport is given by a Wilson line operator 
 \cite{Hansson1988Spin,Grundberg1989, Polyakov1988} 
\beq
\Phi[C] = \langle e^{i \int_{C} A_{\mu}} \rangle = e^{i\theta_{stat} W[C]} = e^{i\theta_{\rm stat}L} e^{-i \theta_{\rm stat} T[C]}
\label{sec2sub1:berryphase}
\eeq
where we have introduced  the writhing number $W[C]$, defined as  $W[C] = L - T[C]$, where $L$ is the linking number and $T[C]$ is the torsion
 \cite{Hansson1988Spin,Grundberg1989, Polyakov1988,Tze1988Manifold} (or the twist) of the curve $C$. Because $L$ is always integer, it is independent of the background metric. The torsion represents how fast the frame of the curve rotates along $C$  
\beq
T[C] = \frac{1}{2\pi} \int d{\bm r} \cdot \left[ {\bm e}_{2} \times \frac{\partial {\bm e}_{2}}{\partial s} \right]
\label{sec2sub1:twist}
\eeq 
We have chosen the frame along the curve to be $\bm{e}_{1} = \frac{\partial {\bm r}}{\partial s}$ and ${\bm e}_{2} \perp {\bm e}_{1}$. When the curvature is purely spatial, and in the absence of  torsion in $\Sigma^{2}$,  we can prove that the phase factor Eq.\eqref{sec2sub1:berryphase} reduces to
\begin{align}
\Phi[C] &= \exp(iL\theta_{\rm stat} ) \exp \left(-iS_{z} \int d{\bm r} \cdot {\bm \omega} \right).
\label{sec2sub1:spinconnection} 
\end{align}

To prove Eq.\eqref{sec2sub1:spinconnection} from Eq.\eqref{sec2sub1:twist} and Eq.\eqref{sec2sub1:berryphase}, we rewrite the torsion as 
\begin{align}
T = \frac{1}{2\pi} \int d{\bm r} \cdot ({\bm e}_{2} \times \frac{\partial {\bm e}_{2}}{\partial s})= \frac{1}{2\pi} \int ds ~\frac{\partial {\bm r}}{\partial s} \cdot ({\bm e}_{2} \times \frac{{\partial \bm e}_{2}}{\partial s}). 
\label{app1:twist1}
\end{align}
We write the vectors appearing in Eq.\eqref{app1:twist1} explicitly. 
\begin{align}
&{\bm r} = (x_{1}(s), x_{2}(s), t(s)), ~ s\in [s_{i}, s_{f}] \nonumber\\
&{\bm e}_{1} = \frac{\partial {\bm r}}{\partial t} = (v_{1}, v_{2}, \alpha), ~ {\bm e}_{2} = \frac{1}{\sqrt{{\bm v}^{2}}}(-v_{2}, v_{1}, 0),\nonumber\\ 
&{\bm v} = (v_{1}, v_{2}, 0)
\label{app1:frame_curve}
\end{align}
Here $s$ is the arc length of $C$, and thus we have taken ${\bm e}_{1} = \frac{\partial {\bm r}}{\partial s}$. It is clear that ${\bm e}_{2} \cdot {\bm e}_{1} =0 $ from the expression. As we are interested in the adiabatic transport of the particle, we impose the condition
\beq
{\bm v}^{2} << \alpha^{2},  
\label{app1:adiabatic}
\eeq 
along the curve $C$. Then this translates as $\alpha = 1 + O(\frac{{\bm v}^{2}}{\alpha^{2}}) $. As the space $\Sigma^{2}$ is curved, we introduce a static local frame on the space. 
\begin{align}
&{\bm E}_{1} = \left(u_{1}(x_{1}, x_{2}),u_{2}(x_{1}, x_{2}),0 \right)\nonumber\\ 
&{\bm E}_{2} = \left(-u_{2}(x_{1}, x_{2}),u_{1}(x_{1}, x_{2}),0 \right)\nonumber\\ 
&{\bm E}_{3} = \left(0,0,1 \right)\nonumber\\ 
\label{app1:frame_space}
\end{align}
Below we will suppress the depedence of $u_{i}$ on $(x_{1}, x_{2}, 0) \in \Sigma^{2}$. However it is important to remember that the frame Eq. \eqref{app1:frame_space} depends on the position because of non-zero curvature in $\Sigma^{2}$. When there is no torsion in $\Sigma^{2}$, the frame follows the equation of motion dictated by the spin connection $\omega_{\mu:ab}$
\beq
\partial_{\mu} E_{a;\nu} - \partial_{\nu}E_{a;\mu} = -\omega_{\mu;ab}E_{b;\nu} +  \omega_{\nu;ab}E_{b;\mu}.
\label{app1:frame_spinconnection}
\eeq
Because the curvature is solely from the space $\Sigma^{2}$, the only non-zero element of the spin connection is $\omega_{\mu:12} = - \omega_{\mu:21} = \omega_{\mu}$ (we suppress the Lorentz indices `$ab$' in the spin connection from here and on). The equation of motion Eq. \eqref{app1:frame_spinconnection} implies that we have the following equations when we are translating the frame along $\frac{\partial {\bm r}}{\partial s}$  
\begin{align}
&\left(\frac{\partial {\bm r}}{\partial s}\cdot {\bm \nabla} \right) {\bm E}_{1} = \left(\frac{\partial {\bm r}}{\partial s}\cdot {\bm \omega} \right)  {\bm E}_{2}, \nonumber\\ 
&\left(\frac{\partial {\bm r}}{\partial s}\cdot {\bm \nabla} \right) {\bm E}_{2} = -\left(\frac{\partial {\bm r}}{\partial s}\cdot {\bm \omega} \right)  {\bm E}_{1}, \nonumber\\ 
\label{app1:spinconnection}
\end{align}
with ${\bm \nabla}$ the covariant derivative. Further we represent ${\bm e}_{2}$ in terms of ${\bm E}_{i}, i =1,2$ by introducing an angle $\phi (s)$. 
\beq
{\bm e}_{2} = \cos(\phi (s)) {\bm E}_{1} + \sin(\phi (s)) {\bm E}_{2}
\eeq
The dependence of $\phi(s)$ on the arc length $s$ represents the relative rotation of the frame ${\bm e}_{i}$ of the curve to the frame ${\bm E}_{i}$ of the space $\Sigma^{2}$. With these in hand, we can proceed to rewrite the twist Eq. \eqref{app1:twist1}, 
\begin{widetext}
\begin{align}
T&= \frac{1}{2\pi} \int ds ~\frac{\partial {\bm r}}{\partial s} \cdot ({\bm e}_{2} \times \frac{{\partial \bm e}_{2}}{\partial s}), \nonumber\\ 
&= \frac{1}{2\pi} \int ds~\frac{\partial {\bm r}}{\partial s} \cdot  \Big\{ \left({\bm E}_{1} \cos\phi  + {\bm E}_{2} \sin\phi  \right) \times \Big[ \frac{\partial \phi}{\partial s}\left( -  {\bm E}_{1} \sin \phi  +{\bm E}_{2} \cos \phi  \right) + \left(\frac{\partial {\bm E}_{1}}{\partial s} \cos\phi  + \frac{\partial {\bm E}_{2}}{\partial s}\sin\phi \right)\Big] \Big\}, \nonumber\\ 
&=\frac{1}{2\pi} \int ds~\frac{\partial {\bm r}}{\partial s} \cdot  \left[ \left({\bm E}_{1} \cos\phi  + {\bm E}_{2} \sin\phi  \right)\times \left( -  {\bm E}_{1} \sin \phi  +{\bm E}_{2} \cos \phi  \right)   \right] \left(\frac{\partial \phi}{\partial s} + \frac{\partial {\bm r}}{\partial s}\cdot {\bm \omega}\right),\nonumber\\ 
&= \frac{1}{2\pi} \int ds~ \left(\frac{\partial {\bm r}}{\partial s} \cdot {\bm E}_{3}\right)  \left(\frac{\partial \phi}{\partial s} + \frac{\partial {\bm r}}{\partial s}\cdot {\bm \omega}\right),\nonumber\\
&=\frac{1}{2\pi} \int ds~ \left(\frac{\partial \phi}{\partial s} + \frac{\partial {\bm r}}{\partial s}\cdot {\bm \omega}\right) + O\left(\frac{{\bm v}^{2}}{\alpha^{2}}\right), \nonumber\\ 
&=\frac{1}{2\pi} \left[ \phi(s_{f}) - \phi(s_{i}) \right] +\frac{1}{2\pi} \int d{\bm r} \cdot {\bm  \omega}. 
\end{align}
\end{widetext}
We have used elementary chain rules and Eq. \eqref{app1:spinconnection} in the second and third lines. Between the fourth line and the fifth line, we have used Eq. \eqref{app1:frame_curve} and Eq. \eqref{app1:frame_space}. Then the last line is just rewriting the integral in a way that it is apparently parameterization independent within the approximation Eq. \eqref{app1:adiabatic} (this approximation becomes exact if the transport is performed {\it infinitely slowly} ${\bm v}^{2} \rightarrow 0$). The first term in the last line is non-universal and depends only on the boundary condition. The term may be dropped out by imposing a periodic boundary condition at $s_{i}$ and $s_{f}$, {\it i.e.,} we may impose that the configuration of the frame at $s = s_{i}$ is the same as that of the frame at $s = s_{f}$. So we drop it in Eq.\eqref{sec2sub1:spinconnection} and from here and on.

Thus the covariant derivative of the composite particle should also include the spin connection with  coupling strength $S_{z}$,
\beq
D_{\mu} = \partial_{\mu} +i a_{\mu}+ iS_{z} \omega_{\mu}
\eeq
This is one of the key results in this paper. Notice that this spin connection is abelian (in contrast to the conventional spin connection of relativistic fermions which is non-abelian.) The composite fermion (CF) and composite boson (CB) CS theories %so far 
in literature are restricted to  flat space, and hence there is no need to introduce the spin connection explicitly. However,  the geometric response involves the deformation of the metric, and it is necessary to keep the spin connection explicitly. We will show that inclusion of the spin connection in the covariant derivative leads to the correct Hall viscosity and Wen-Zee term for abelian FQH fluids. 

\section{Geometry in the Composite Fermion theory}
\label{sec:CF}

\subsection{Laughlin and Jain States}
\label{sec:laughlin-jain}

We first consider the CF theory \cite{Lopez1991, Jain1989} of a FQH state at the filling $\nu = \frac{1}{2p+1}$ in a curved space. We begin with the action of the non-relativistic Fermi field  $\Psi_{e}$ describing the dynamics of electrons in two dimensions under an uniform magnetic field,
\begin{align}
S&=\int d^3x  \sqrt{g} \bigg[ \frac{i}{2}\left( \left( D_{0} \Psi_{e}(x) \right)^{\dagger} \Psi_{e} - \Psi^{\dagger}_{e}(x) \left( D_{0} \Psi_{e}(x) \right)  \right) \nonumber\\
 & - \frac{1}{2} (D_{i}\Psi_{e} (x))^{\dagger}g^{ij}(D_{j}\Psi_{e} (x) ) \bigg] + S_{\rm int}, 
\label{sec2sub2:original}
\end{align}
 in which $D_{\mu} = \partial_{\mu} + iA_{\mu}$ is the covariant derivative of the electron, and we set the effective mass $m_e$ and the charge of electron to be unity. The electron is a scalar field, and thus does not couple minimally with the spin connection. $S_{\rm int}$ encodes the short-ranged repulsive density-density interaction between electrons. The interaction term will not affect the Hall viscosity \cite{You_unpublished}, and so it can be ignored from here and on. The electromagnetic gauge field $A_{\mu}$ can be written as $A_{\mu} = {\bar A}_{\mu} + \delta A_{\mu}$ where ${\bar A}_{\mu}$ is  the uniform magnetic field, and $\delta A_{\mu}$ is a probe field that measures the electromagnetic response of the FQH state.
 % We are interested in the linear response of the FQH state to the shear metric deformation $\delta g_{ij} = g_{ij} - \delta_{ij}$ 
 %such as Tr $(\delta g_{ij}) =0$ and $g = \det g_{ij} = 1 + O(\delta g^{2}_{ij})$. 

The fermion Chern-Simons field theory of the FQH states \cite{Lopez1991,fradkin-1991} consists of attaching an even number of flux quanta to each electron by formally coupling the theory of Eq.\eqref{sec2sub2:original} to an abelian (statistical) gauge field $a_\mu$ whose Lagrangian has the CS form. The resulting action in terms of the composite fermion $\Psi$ is
\begin{align} 
S&=\int d^3x  \sqrt{g} \Big[\frac{i}{2}\left( \left( D_{0} \Psi(x) \right)^{\dagger} \Psi - \Psi^{\dagger}(x) \left( D_{0} \Psi(x) \right)  \right) \nonumber\\
& - \frac{1}{2} (D_{i}\Psi (x))^{\dagger}g^{ij}(D_{j}\Psi (x) )   + \frac{\varepsilon^{\mu\nu\lambda}}{8\pi p}a_{\mu}\partial_{\nu}a_{\lambda} \Big]
\end{align}
where $D_{\mu} = \partial_{\mu} + iA_{\mu} + ia_{\mu} + ip \omega_{\mu}$ is a covariant derivative which, in addition to the minimal coupling to the statistical gauge field $a_\mu$, includes the minimal coupling to the spin connection $\omega_\mu$. The CS term binds the $2p$ flux quanta to the electron $\Psi_{e}$ and turns the electron into the CF $\Psi$.\cite{Jain1989} Notice that  the spin connection  enters explicitly in the covariant derivative with a topological spin $p \in {\mathbb Z}$ reflecting the statistical angle $\theta_{\rm stat} = 2\pi p$. 

The FQH state is described in this CF picture \cite{Jain1989,Lopez1991} by noting that if we attached $2p$ flux quanta to each fermion, on average the external vector potential ${\bar A}_j$ is partially screened to ${\bar A}_{j} + {\bar a}_{j} = \frac{1}{2p+1} {\bar A}_{j}$. Thus the CF $\Psi$ is subject to the magnetic field which is $\frac{1}{2p+1}$ of the magnetic field experienced by the bare electron. Then for a system with filling $\nu=\frac{1}{2p+1}$ the CF fill up  the lowest effective Landau level.\cite{Jain1989} The  FQH effect is then obtained by integrating out the CF fluctuations at the one-loop level. 

Next we note that integrating out the CF fluctuations at $\nu = \frac{1}{2p+1}$ is formally equivalent to integrating out the electron fluctuations in the integer quantum Hall fluid at $\nu =1$, {\it i.e.,} the theory Eq.\eqref{sec2sub2:original} at the filling $\nu =1$, which is done in  
Ref.[\onlinecite{Gromov2014}] and Ref.[\onlinecite{ You_unpublished}]. For the integer quantum Hall state, we obtain the effective theory $\mathcal{L}=\mathcal{L}_0+\mathcal{L}_{\rm topo}$ of the electromagnetic probe $\delta A_\mu$, and of the metric $\delta g_{ij}$,
\begin{align} 
&{\mathcal L}_{0} = \delta A_0 {\bar \rho}  + {\bar \rho} s_{0} \omega_0,\\ 
&{\mathcal L}_{\rm topo} =  \varepsilon^{\mu\nu\lambda} \bigg[ \frac{1}{4\pi} \delta A_{\mu}\partial_{\nu} \delta A_{\lambda}  +  \frac{s_{0}}{2\pi} \delta A_{\mu}\partial_{\nu} \omega_{\lambda} + \frac{1}{24\pi} \omega_{\mu}\partial_{\nu} \omega_{\lambda} \bigg] 
\label{sec2sub2:integer}
\end{align}
in which $s_{0} = \frac{1}{2}$, the average orbital spin of the integer quantum Hall state. Here we have $\omega_{i} = - \frac{1}{2}\varepsilon^{jk} \partial_{j}\delta g_{ik}$ and $\omega_{t} = \frac{1}{2}\varepsilon^{jk}\delta g_{ij}\partial_{t} \delta g_{ik}$.\cite{Gromov2014, Gromov2014b, Son2013} The second term in ${\mathcal L}_{0}$ is the Berry phase term of the effective action which accounts for the Hall viscosity, and the second term in ${\mathcal L}_{\rm topo}$ is the Wen-Zee term. The last term in ${\mathcal L}_{\rm topo}$ is the gravitational CS term\cite{Gromov2014} (see below).

Having the effective theory Eq.\eqref{sec2sub2:integer} in hand, we can easily obtain the effective theory of the fluctuating component  $\delta a_\mu$  of the statistical gauge field, of the electromagnetic probe $\delta A_\mu$, and of the metric $\delta g_{ij}$ in the FQH state from Eq.\eqref{sec2sub2:integer}. We replace $\delta A_{\mu}$ in Eq.\eqref{sec2sub2:integer} to $\delta A_{\mu} + \delta a_{\mu} + p \omega_{\mu}$ to obtain that of the FQH state because the CF field minimally couples to $\delta A_{\mu} + \delta a_{\mu} + p \omega_{\mu}$. Then the resulting effective Lagrangian to lowest orders in a gradient expansion again has the form $\mathcal{L}=\mathcal{L}_0+\mathcal{L}_{\rm topo}$ where \cite{You_unpublished,Gromov2014} 
\begin{align} 
 {\mathcal L}_{0} &= \left( \delta A_0 +  p \omega_0 \right) {\bar \rho}  + {\bar \rho} s_{0} \omega_0,\nonumber\\
 {\mathcal L}_{\rm topo} &=  \frac{\varepsilon^{\mu\nu\lambda}}{4\pi} (\delta A_{\mu} + \delta a_{\mu} + p \omega_{\mu})\partial_{\nu} (\delta A_{\lambda} + \delta a_{\lambda} + p \omega_{\lambda}) \nonumber\\ 
 &+ \frac{\varepsilon^{\mu\nu\lambda}}{4\pi} (\delta A_{\mu} + \delta a_{\mu} + p \omega_{\mu})\partial_{\nu} \omega_{\lambda} + \frac{\varepsilon^{\mu\nu\lambda}}{24\pi} \omega_{\mu}\partial_{\nu} \omega_{\lambda} \nonumber\\
&+ \frac{\varepsilon^{\mu\nu\lambda}}{8\pi p}\delta a_{\mu}\partial_{\nu} \delta a_{\lambda},
\label{sec2sub2:intermediate}
\end{align}
where $s_{0} =1/2$ is the orbital spin of a system of fermions at $\nu =1$, and ${\bar \rho}$ is the electron density. The last term in ${\mathcal L}_{\rm topo}$, the CS term of $\delta a_{\mu}$, comes from the CS term responsible for flux attachment. 

In the above discussion, we did not include explicitly the short-ranged repulsive density-density interaction. Although the interactions are obviously necessary to stabilize the FQH state, in the effective action of the excitations  their contribution  interaction enters only in  the Maxwell term of the statistical gauge field, which is non-topological and subleading to the Chern-Simons terms. Thus  the universal geometric response does not depend explicitly in the form of the interactions.

\subsubsection{Hydrodynamic theory: Hall viscosity and Wen-Zee term}

We can further  transform the effective theory of Eq.\eqref{sec2sub2:intermediate} into the hydrodynamic theory of the FQH state \cite{Wen95,wenrev1}.
To this end we introduce the hydrodynamic field $b_{\mu}$, rewrite the last term in Eq.\eqref{sec2sub2:intermediate} as
\begin{align}
\frac{\varepsilon^{\mu\nu\lambda}}{8\pi p}\delta a_{\mu}\partial_{\nu} \delta a_{\lambda} \rightarrow - \frac{2p}{4\pi}\varepsilon^{\mu\nu\lambda} b_{\mu} \partial_{\nu}b_{\lambda} + \frac{\varepsilon^{\mu\nu\lambda}}{2\pi}b_{\mu}\partial_{\nu} \delta a_{\lambda},
\end{align}
and integrate-out the fluctuation $\delta a_{\mu}$ from Eq.\eqref{sec2sub2:intermediate} to obtain the hydrodynamic theory for the FQH state   
 \cite{Wen95,wenrev1}, which now also includes the coupling to a curved space
\begin{align}
\mathcal{L} &= +{\bar \rho} \delta A_0 + \frac{2p+1}{2} {\bar \rho}\omega_0 -\frac{2p+1}{4\pi} \varepsilon^{\mu\nu\lambda} b_{\mu}\partial_{\nu} b_{\lambda} \nonumber\\
&-\frac{\varepsilon^{\mu\nu\lambda}}{2\pi} b_{\mu}\partial_{\nu}\delta A_{\lambda} - \frac{2p+1}{2} \frac{\varepsilon^{\mu\nu\lambda}}{2\pi} b_{\mu}\partial_{\nu} \omega_{\lambda}  - \frac{\varepsilon^{\mu\nu\lambda}}{48\pi} \omega_{\mu}\partial_{\nu}\omega_{\lambda} 
\label{sec2sub2:hydrodynamic}
\end{align}
This effective Lagrangian has the same form as the hydrodynamic theory \cite{Wen95}. However,  Eq.\eqref{sec2sub2:hydrodynamic} also includes the Berry phase term of the Hall viscosity (second term of the first line), and the Wen-Zee term \cite{wenzee} (second term  of the second line). The last term has the form of gravitational CS terms \cite{Deser-1982} and will be discussed below. 
 
The Hall viscosity of the FQH state is obtained by varying ${\mathcal L}$ in Eq.\eqref{sec2sub2:hydrodynamic} with respect to the metric  $\delta g_{ij}$. We find
\beq
\eta_{H} = s \frac{{\bar \rho}}{2} = \frac{2p+1}{2}\frac{{\bar \rho}}{2},  
\eeq
which  agrees with previous results obtained by other arguments.\cite{Read2009,Read2011,Hansson2013,Bradlyn2012}  Here $s$ is the intrinsic orbital spin, $s=\frac{2p+1}{2}$.

\subsubsection{The gravitational Chern-Simons term}

We can further identify the 
%chiral 
central charge of the edge states of this FQH fluid  by reading-off the coefficient  of the gravitational CS term, $-\frac{c}{48\pi} \varepsilon^{\mu\nu\lambda}\omega_{\mu} \partial_{\nu} \omega_{\lambda} $, in Eq.\eqref{sec2sub2:hydrodynamic} and find $c=1$. The gravitational CS term embodies the gravitational anomaly of the energy-momentum tensor in topological fluids.\cite{Ryu-2012,Stone-2012,Luttinger-1964}
% A standard result from quantum field theory (see, e.g. Ref. \cite{weinberg}) relates the 
 The response to the effective action to a change of the metric (and hence of the spin connection) to the expectation value of the energy-momentum 
 tensor.\cite{weinberg}
  % of these systems. in particular, 
  Given the (holographic) correspondence 
  %that exists 
  between the bulk and the edge states of quantum Hall fluids,\cite{Wen95} 
  (reflecting the holographic nature of Chern-Simons gauge theory\cite{Witten1989,Elitzur1989})
  %it is expected that 
  the central charge derived from the gravitational CS term should be the same as the central charge of the theory of the edge states, %
  %and hence to be 
  determined by the level of the CS term for the hydrodynamic gauge fields.

At this stage, one might conclude that the CF theory can be used to predict the correct central charge of the FQH state. However, this is the {\it artifact} of the mean-field state that we choose to study. To see this, we consider another legitimate CF construction for the Laughlin state at $\nu = \frac{1}{2p+1}$ in which we attach ($2p+2$) flux quanta to electron $\Psi_{e}$ in Eq.\eqref{sec2sub2:original}. 
%\begin{align} 
%S&=\int d^3x  \sqrt{g} \Big[\frac{i}{2}\left( \left( D_{0} \Psi(x) \right)^{\dagger} \Psi - \Psi^{\dagger}(x) \left( D_{0} \Psi(x) \right)  \right) \nonumber\\
%& - \frac{1}{2} (D_{i}\Psi (x))^{\dagger}g^{ij}(D_{j}\Psi (x) )   + \frac{\varepsilon^{\mu\nu\lambda}}{8\pi (p+1)}a_{\mu}\partial_{\nu}a_{\lambda} \Big]
%\label{CF:artifact}
%\end{align}
%where $D_{\mu} = \partial_{\mu} + iA_{\mu} + ia_{\mu} + i(p+1)\omega_{\mu}$ is a covariant derivative. 
Then the resulting CF %$\Psi$ 
sees {\it on average} one flux quanta, which is directed opposite from the direction of the external magnetic flux, per particle. Thus we choose the mean-field state in which the CF is in the $\nu=-1$ state, in which the CF fills up the lowest Landau level, {\it i.e.,} ${\bar A}_{j} + {\bar a}_{j} = -\frac{1}{2p+1} {\bar A}_{j}, j =x,y$. The state has the opposite chirality from the integer quantum Hall state at $\nu =1$. 

We can now follow the same steps that we used to derive the effective hydrodynamic theory. The resulting effective action , i.e. the analog of Eq.\eqref{sec2sub2:hydrodynamic}, now is
\begin{align}
\mathcal{L} &= +{\bar \rho} \delta A_0 + \frac{2p+1}{2} {\bar \rho}\omega_0 -\frac{2p+1}{4\pi} \varepsilon^{\mu\nu\lambda} b_{\mu}\partial_{\nu} b_{\lambda} \nonumber\\
&-\frac{\varepsilon^{\mu\nu\lambda}}{2\pi} b_{\mu}\partial_{\nu}\delta A_{\lambda} - \frac{2p+1}{2} \frac{\varepsilon^{\mu\nu\lambda}}{2\pi} b_{\mu}\partial_{\nu} \omega_{\lambda}  + \frac{\varepsilon^{\mu\nu\lambda}}{48\pi} \omega_{\mu}\partial_{\nu}\omega_{\lambda}.
\label{sec2sub2:hydrodynamic2}
\end{align}
Compared to Eq.\eqref{sec2sub2:hydrodynamic}, we see that only the last term has the {\it wrong} coefficient $c= -1$, which is the central charge of the mean-field state! This implies that the correct central charge in Eq. \eqref{sec2sub2:hydrodynamic} is an artifact of the mean-field state which accidentally has the same central charge as the Laughlin state. Except the coefficient of the gravitational CS term, {\it i.e.,} central charge, the Hall viscosity and the Wen-Zee term are, here too, correctly reproduced and are independent of the mean-field state of the CF theory. We will see that the CB theory and the projective parton constructions suffer from the same problem and the predict the  central charge to be that of the mean-field states, which in general is not  the correct central charge of the system. 

%This analysis generalizes to multi-component abelian states as done below.
  %but we will see below that there are some open issues in the 
%  and non-abelian states.

\subsection{Multi-component abelian FQH States}
\label{sec:multi-abelian}

Having the descriptions Eq.\eqref{sec2sub2:hydrodynamic} and Eq.\eqref{sec2sub2:intermediate} of the Laughlin states in hand, we can proceed to study the bilayer quantum Hall state from the composite fermion theory (where repeated indices are summed over)
\begin{align}
{\mathcal L} &= \sqrt{g} \bigg[ \frac{i}{2}\left( \left( D^{a}_{0} \Psi_{a}(x) \right)^{\dagger} \Psi_{a} - \Psi^{\dagger}_{a}(x) \left( D^{a}_{0} \Psi_{a}(x) \right)  \right)  \nonumber\\ 
& - \frac{1}{2} (D^{a}_{i}\Psi_{a} (x))^{\dagger}g^{ij}(D^{a}_{j}\Psi_{a} (x) ) \bigg] + {\mathcal L}_{CS}, 
\label{sec2sub2:Bilayeroriginal}
\end{align}
 in which $D^{a}_{\mu} = \partial_{\mu} + iA^{a}_{\mu}+i\alpha^{a}_{\mu} + ip_{a}\omega_{\mu}$ is the covariant derivative of the composite fermions in the layer $a=1,2$. As in the single layer case, the electrons are attached with fluxes of the statistical gauge fields $\alpha^{a}_{\mu}$ and turned into the composite fermions.  Here too, to simplify the notation, we do not include the interaction terms explicitly since they do not affect the topological structure. Naturally, the interactions  are crucial to stabilize the FQH state and these particles are not free but strongly interacting.
 
 The CS term ${\mathcal L}_{CS}$ in \eqref{sec2sub2:Bilayeroriginal} is 
\begin{align}
&\mathcal{L}_{CS}=\frac{\varepsilon^{\mu\nu\lambda}}{4\pi} \alpha^{a}_{\mu} \left[ K^{-1}\right]^{ab} \partial_{\nu}\alpha^{b}_{\lambda}, ~ K = 
\begin{pmatrix} 
2p_{1} & n  \\
n & 2p_{2}
\end{pmatrix},
\end{align}
and where $a,b=1,2$, and $p_1$, $p_2$ and $n$ are arbitrary integers. Notice that the spins of the composite fermions depend only on the change in the self-statistical angle $\theta_{a} = 2\pi p_{a}$ and thus the fermions couple to the spin connection with the strength of $p_{a}$, {\it i.e.,} the diagonal elements of $K$-matrix in ${\mathcal L}_{CS}$.

We smear out the fluxes of the statistical gauge fields into space and assume that the composite fermions are at $\nu^{a}=1, a= 1,2$ (the generalization to the other values of $\nu^{a} \in {\mathbb Z}$ is straightforward). We integrate out the composite fermions and expand the resulting effective Lagrangian ${\mathcal L}$ in terms of the perturbations $\{ \delta \alpha^{a}_{\mu}, \delta A^{a}_{\mu}, \delta g^{ij} \}$ around their mean values 
\begin{align}
{\mathcal L} =&{\mathcal L}_{0} + {\mathcal L}_{topo} + \cdots, \nonumber\\ 
{\mathcal L}_{0} =&  (\delta A^{a} + p_{a} \omega_{t}) {\bar \rho}^{a} + \frac{1}{2} {\bar \rho}^{a} \omega_{t} ,  \nonumber\\ 
{\mathcal L}_{\rm topo} =&   \frac{\varepsilon^{\mu\nu\lambda}}{4\pi} \delta \kappa^{a}_{\mu} \partial_{\nu} \delta \kappa^{a}_{\mu}   + \frac{\varepsilon^{\mu\nu\lambda}}{4\pi}\delta \kappa^{a}_{\mu} \partial_{\nu} \omega_{\lambda} + \frac{\varepsilon^{\mu\nu\lambda}}{24\pi} \omega_{\mu}\partial_{\nu} \omega_{\lambda}  \nonumber\\
 + &\frac{\varepsilon^{\mu\nu\lambda}}{4\pi} \delta \alpha^{a}_{\mu} \left[ K^{-1}\right]^{ab} \partial_{\nu}\delta \alpha^{b}_{\lambda}, 
\label{sec2sub2:Bilayerintermediate}
\end{align}   
in which $\delta \kappa^{a}_{\mu} = \delta A^{a}_{\mu} + \delta \alpha^{a}_{\mu} + p_{a} \omega_{\mu}$. We can transform this effective theory Eq. \eqref{sec2sub2:Bilayerintermediate} into the hydrodynamic description by rewritting the last term of ${\mathcal L}_{topo}$ in Eq. \eqref{sec2sub2:Bilayerintermediate} in terms of the hydrodynamics fields $\beta^{a}_{\mu}$. 
\begin{align}
{\mathcal L}_{CS} =& \frac{\varepsilon^{\mu\nu\lambda}}{4\pi} \delta \alpha^{a}_{\mu} \left[ K^{-1}\right]^{ab} \partial_{\nu}\delta \alpha^{b}_{\lambda} \nonumber\\
  \rightarrow& -\frac{\varepsilon^{\mu\nu\lambda}}{4\pi}  \beta^{a}_{\mu}  K^{ab} \partial_{\nu}\beta^{b}_{\lambda}+\frac{\varepsilon^{\mu\nu\lambda}}{2\pi}  \beta^{a}_{\mu} \partial_{\nu}\delta \alpha^{a}_{\lambda}
\end{align}
With this result  in hand, we integrate out $\delta \alpha^{a}_{\mu}$ from \eqref{sec2sub2:Bilayerintermediate} to find 
\begin{align}
{\mathcal L} =&{\mathcal L}_{0} + {\mathcal L}_{topo} + \cdots, \nonumber\\ 
{\mathcal L}_{0} =& {\bar \rho}^{a} \delta A^{a}_{t} +(p_{a}+ \frac{1}{2}) {\bar \rho}^{a} \omega_{t} , \nonumber\\ 
{\mathcal L}_{topo} =& -\frac{\varepsilon^{\mu\nu\lambda}}{4\pi}  \beta^{a}_{\mu}  {\widetilde K}^{ab} \partial_{\nu}\beta^{b}_{\lambda} \nonumber\\
 +& -\frac{\varepsilon^{\mu\nu\lambda}}{2\pi} \beta^{a}_{\mu}\partial_{\nu} \delta A^{a}_{\lambda} - \frac{ p_{a}+\frac{1}{2}}{2\pi} \varepsilon^{\mu\nu\lambda} \beta^{a}_{\mu}\partial_{\nu} \omega_{\lambda} \nonumber\\
 - & \frac{\varepsilon^{\mu\nu\lambda}}{48\pi} \omega_{\mu}\partial_{\nu} \omega_{\lambda}, 
\label{sec2sub2:Bilayerfinal}
\end{align}
where
\begin{align}
&{\widetilde K} = 
\begin{pmatrix} 
2p_{1} +1& n  \\
n & 2p_{2}+1
\end{pmatrix}.
\end{align}   
The Hall viscosity of this bilayer system is,
\beq
\eta_{H} = \sum_{a=1,2} (p_{a}+ \frac{1}{2}) \frac{{\bar \rho}^{a}}{2}, 
\eeq
in agreement with Refs.[\onlinecite{Hansson2013}] and  [\onlinecite{Gromov2014b}]. Furthermore, these results yield the correct value of the  Wen-Zee term in Eq. \eqref{sec2sub2:Bilayerfinal}  with the correct spin for the bilayer system.\cite{wenzee, Wen95} Thus, as  in the CF description of the Laughlin states, the Hall viscosity and the Wen-Zee term are however correctly reproduced, independent of the choice of the mean-field states. Finally, it is straightforward to generalize the theory present here to any abelian multi-component FQH states. 

Finally, here too, from the coefficient of the gravitational CS term $-\frac{c}{48\pi}\varepsilon^{\mu\nu\lambda}\omega_{\mu}\partial_{\nu}\omega_{\lambda}$ in \eqref{sec2sub2:Bilayerfinal}, we infer that the chiral central charge of the theory is $c=2$. However, just as in the  case of the Laughlin and Jain states, this result  is again an artifact of the mean-field states, which accidentally have the same central charge as the physical states that we are studying. 

\section{Geometry in the Composite Boson theory} 
\label{sec:CB}

\subsection{Laughlin States}
\label{sec:laughlin-cb}

 %It is instructive to
We can do the same analysis in the CB theory,\cite{Zhang1989,Zhang1992} again for  a FQH state with filling fraction $\nu=1/(2p+1)$. The main difference with the CF theory is that the theory of fermions in a magnetic field is now mapped onto a system of with a Bose  field $\Phi$ coupled to the CS theory% with the action
\begin{align}
S &=\int d^3x  \sqrt{g} \bigg[ \frac{i}{2}\left( \left( D_0 \Phi(x) \right)^{\dagger} \Phi  - \Phi^{\dagger}(x) \left( D_0 \Phi(x) \right) \right) \nonumber\\
 &- \frac{1}{2} (D_i\Phi (x))^{\dagger}g^{ij}(D_j\Phi (x) ) + \frac{\varepsilon^{\mu\nu\lambda}}{4\pi (2p+1)} a_{\mu}\partial_{\nu}a_{\lambda} \bigg], 
\label{sec2sub3:original}
\end{align}
 in which $D_{\mu} = \partial_{\mu} + iA_{\mu}+ia_{\mu}+ i\frac{2p+1}{2}\omega_{\mu}$ is the covariant derivative of the CB on a curved manifold.\cite{Son2013} We can perform the standard dual transformation\cite{Zhang1992} of the CB theory on the action Eq.\eqref{sec2sub3:original}. We start by rewriting Eq.\eqref{sec2sub3:original} as 
\begin{align}
{\mathcal L} &= \sqrt{g} \bigg[ \frac{i}{2}\left( \left( D_{0} \Phi(x) \right)^{\dagger} \Phi  - \Phi^{\dagger}(x) \left( D_{0} \Phi(x) \right) \right) \nonumber\\
& - \frac{1}{2} (D_{i}\Phi (x))^{\dagger}g^{ij}(D_{j}\Phi (x) ) + \frac{\varepsilon^{\mu\nu\lambda}}{4\pi (2p+1)} a_{\mu}\nabla_{\nu}a_{\lambda} \bigg]. 
\end{align}
Here the CS term is written in a way that it is explicitly invariant under general coordinate transformation by using the covariant derivative $\nabla_{\nu}$   
\beq
\varepsilon^{\mu\nu\lambda} \nabla_{\nu} a_{\lambda} = \varepsilon^{\mu\nu\lambda}\bigg( \partial_{\nu}a_{\lambda} + \Gamma^{\sigma}_{\nu\lambda}a_{\sigma}\bigg) = \varepsilon^{\mu\nu\lambda}\partial_{\nu}a_{\lambda},
\eeq
because of the property of the Christoffel symbol, $\Gamma^{\sigma}_{\nu\lambda} = \Gamma^{\sigma}_{\lambda\nu}$. The Levi-Civita tensor is normalized as $\varepsilon^{txy} = \frac{1}{\sqrt{g}}$. In the composite boson theory, the FQH state of the electron corresponds to the superfluid state of the boson $\Phi$. In the superfluid phase, the average ${\bar A}_{\mu}$ of the electromagnetic gauge field $A_{\mu} = {\bar A}_{\mu} + \delta A_{\mu}$ is completely cancelled by the average ${\bar a}_{\mu}$ of the statistical gauge field $a_{\mu} = {\bar a}_{\mu} + \delta a_{\mu}$, {\it i.e.,} ${\bar A}_{i} + {\bar a}_{i}= 0, i =x, y$. On the other hand, the average density of the boson is locked with the average magnetic field due to the quantum Hall effect. 
\beq
\langle \Phi^{\dagger}\Phi \rangle = {\bar \rho} = \frac{1}{2\pi k } \varepsilon^{t ij} \nabla_{i}{\bar A}_{j} = -\frac{1}{2\pi k } \varepsilon^{t ij} \nabla_{i}{\bar a}_{j} 
\eeq
We can write down the low-energy Lagrangian for the superfluid by expanding $\Phi = \sqrt{{\bar \rho} + \delta \rho} e^{i\theta}$ in terms of $\delta \rho$ and $\theta$,  
\begin{align}
&{\mathcal L} = \sqrt{g}\bigg[ (\partial_{t} \theta + \delta \alpha_{t}){\bar \rho}   + (\partial_{t} \theta + \delta \alpha_{t}+\delta a_{t})\delta \rho \nonumber\\
&\quad - \frac{{\bar \rho}g^{ij}}{2} (\partial_{i} \theta + \delta \alpha_{i} +\delta a_{i}) (\partial_{j} \theta + \delta \alpha_{j}+\delta a_{j})  \nonumber\\ 
&\quad + \frac{\varepsilon^{\mu\nu\lambda}}{4\pi (2p+1)} \delta a_{\mu}\nabla_{\nu} \delta a_{\lambda} \bigg], 
\end{align}
with $\delta \alpha_{\mu} = \delta A_{\mu} + \frac{2p+1}{2}\omega_{\mu}$. Here, the first term $\sim \sqrt{g} {\bar \rho} \partial_{t}\theta$ in the right hand side can be gauged away. We introduce a Hubbard-Stratonovich field $J^{i}$ to rewrite the kinetic term of the composite boson,
\begin{align}
&\sqrt{g} \frac{{\bar \rho}g^{ij}}{2} (\partial_{i} \theta + \delta \alpha_{i} +\delta a_{i}) (\partial_{j} \theta + \delta \alpha_{j}+\delta a_{j}) \nonumber\\ 
&\quad \rightarrow \sqrt{g}\bigg[ (\partial_{i} \theta + \delta \alpha_{i} +\delta a_{i})g^{ij}J_{j} - \frac{1}{2{\bar \rho}} J_{i} g^{ij}J_{j} \bigg].
\end{align}
With these in hand, we have  
\begin{align}
{\mathcal L} =& \sqrt{g}\bigg[ {\bar \rho} \delta \alpha_{t}  +  (\partial_{\mu} \theta + \delta \alpha_{\mu}+\delta a_{\mu})J^{\mu} +\frac{1}{2{\bar \rho}} J_{i} g^{ij}J_{j} \bigg] \nonumber\\ 
&+ \sqrt{g} \frac{\varepsilon^{\mu\nu\lambda}}{4\pi (2p+1)} \delta a_{\mu}\nabla_{\nu} \delta a_{\lambda},
\label{CB:duality}
\end{align}
where $J^{\mu} = (\delta \rho, -J^{i})$ represents the conserved boson current. In the absence of the vortex excitation, we can integrate out the phase variable $\theta \in {\mathbb R}$ to obtain, (but inclusion of the vortex can be done easily)
\beq
\partial_{\mu} (\sqrt{g}J^{\mu}) = \sqrt{g} \nabla_{\mu}J^{\mu} =0 \rightarrow J^{\mu} = \varepsilon^{\mu\nu\lambda} \frac{1}{2\pi} \nabla_{\nu}b_{\lambda}, 
\eeq
in which a hydrodynamic (gauge) field $b_{\mu}$ is introduced to solve the equation of motion. By plugging this back to the Lagrangian Eq. \eqref{CB:duality}, we find  
\begin{align}
&{\mathcal L}  =\sqrt{g} \bigg[ {\bar \rho} \delta \alpha_{t} + \frac{1}{2\pi} \varepsilon^{\mu\nu\lambda} (\delta \alpha_{\mu} + \delta a_{\mu})\nabla_{\nu}b_{\lambda} \nonumber\\ 
&\quad +\frac{\varepsilon^{\mu\nu\lambda}}{4\pi (2p+1)} \delta a_{\mu}\nabla_{\nu} \delta a_{\lambda} - \frac{1}{2{\bar \rho}} e_{i} g^{ij}e_{j} \bigg]. 
\end{align}
Here $e_{i} = \frac{1}{2\pi} \varepsilon^{i\sigma\lambda}\nabla_{\sigma}b_{\lambda}, i=x,y$ is the electric field of $b_{\mu}$. We integrate out $\delta a_{\mu}$ and obtain the effective action for the FQH state in the curved space
\begin{align} 
{\mathcal L}  =& \sqrt{g} \bigg[ {\bar \rho} \delta \alpha_{t} + \frac{1}{2\pi} \varepsilon^{\mu\nu\lambda} \delta \alpha_{\mu} \nabla_{\nu}b_{\lambda} -\frac{2p+1}{4\pi} \varepsilon^{\mu\nu\lambda} b_{\mu}\nabla_{\nu} b_{\lambda} \nonumber\\ 
&- \frac{1}{2{\bar \rho}} e_{i} g^{ij}e_{j} \bigg].
\end{align}
Expanding this effective theory to the leading order of $\delta g_{ij}$ and gauge fields, we find 
\begin{align}
{\mathcal L} = &  \delta A_0  \bar \rho + \frac{2p+1}{2}\omega_0 {\bar \rho} -\frac{2p+1}{4\pi} \varepsilon^{\mu\nu\lambda} b_{\mu}\partial_{\nu} b_{\lambda} \nonumber\\ 
&+  \frac{1}{2\pi} \varepsilon^{\mu\nu\lambda} \delta A_{\mu} \partial_{\nu}b_{\lambda} + \frac{2p+1}{4\pi} \varepsilon^{\mu\nu\lambda} \omega_{\mu} \partial_{\nu}b_{\lambda} +\cdots
\end{align}
The Hall viscosity and the Wen-Zee term (with the correct orbital spin) of the FQH state are correctly reproduced here. Finally, in our analysis of the composite boson theory we did not include explicitly the short-ranged repulsive density-density interaction just as in the composite fermion picture and for the same reasons.
%, but the interaction is turned into the Maxwell term after the dual transformation and is subleading than the CS terms. 
Here too, the interaction, which is crucial for the existence of the FQH states,  does not affect the value of the Hall viscosity and of the Wen-Zee term. 

However, the expected gravitational CS term\cite{Gromov2014, Gromov2014b} is apparently absent in the boson theory. Naively this happens  because the mean-field state of the CB theory is the time-reversal-invariant superfluid phase. Time-reversal symmetry is unbroken at the mean-field level since the external magnetic field is exactly cancelled by the flux of the statistical gauge field. In this picture, the breaking of time reversal invariance enters only through the Chern-Simons term in the effective action.

\subsection{Multi-component FQH states}
\label{sec:multi-cb}

As in the CF theory case, we can proceed to describe the bilayer FQH state by the composite boson theory. We have the two species of the Bose fields $\Phi_{a}, a =1,2$  (again with repeated indices being summed over)
\begin{align}
{\mathcal L} &= \sqrt{g}  \frac{i}{2}\left( \left( D^{a}_{0} \Phi_{a}(x) \right)^{\dagger} \Phi_{a} - \Phi^{\dagger}_{a}(x) \left( D^{a}_{0} \Phi_{a}(x) \right)  \right) \nonumber\\ 
&- \frac{1}{2} (D^{a}_{i}\Phi_{a} (x))^{\dagger}g^{ij}(D^{a}_{j}\Phi_{a} (x) )   + {\mathcal L}_{CS}, 
\label{sec2sub2:Bilayeroriginal:CB}
\end{align}
 in which $D^{a}_{\mu} = \partial_{\mu} + iA^{a}_{\mu}+i\alpha^{a}_{\mu} + i(p_{a}+\frac{1}{2})\omega_{\mu}$ is the covariant derivative of the composite bosons $\Phi_{a}$ in the layer $a=1,2$. The CS term ${\mathcal L}_{CS}$ in \eqref{sec2sub2:Bilayeroriginal:CB} is  
\begin{align}
&\mathcal{L}_{CS}=\frac{\varepsilon^{\mu\nu\lambda}}{4\pi} \alpha^{a}_{\mu} \left[ K^{-1}\right]^{ab} \partial_{\nu}\alpha^{b}_{\lambda},~\nonumber\\
& ~ K=
\begin{pmatrix} 
2p_{1}+1 & n  \\
n & 2p_{2}+1
\end{pmatrix},
\label{CB:CS}
\end{align}
where $a,b=1,2$ and $p_1$, $p_2$ and $n$ are arbitrary integers. Notice that the spins of the composite bosons depend only on the change in the self-statistical angle $\theta_{a} = 2\pi (p_{a}+\frac{1}{2})$ and thus the bosons couple to the spin connection with the strength of $p_{a}+\frac{1}{2}$, {\it i.e.,} the diagonal elements of $K$-matrix in ${\mathcal L}_{CS}$. Then, the FQH state corresponds to the superfluid state of the boson $\Phi_a, a =1,2$. By performing the dual transformation, we find  
 \begin{align}
{\mathcal L} =&{\mathcal L}_{0} + {\mathcal L}_{topo} + \cdots, \nonumber\\ 
{\mathcal L}_{0} =&   {\bar \rho}^{a} \delta A^{a}_{t} +(p_{a}+ \frac{1}{2}) {\bar \rho}^{a} \omega_{t}, \nonumber\\ 
{\mathcal L}_{\rm topo} =& -\frac{\varepsilon^{\mu\nu\lambda}}{4\pi}  \beta^{a}_{\mu} K ^{ab} \partial_{\nu}\beta^{b}_{\lambda}  \nonumber\\ 
+&  \frac{\varepsilon^{\mu\nu\lambda}}{2\pi} \beta^{a}_{\mu}\partial_{\nu} \delta A^{a}_{\lambda} + \frac{ p_{a}+\frac{1}{2}}{2\pi} \varepsilon^{\mu\nu\lambda} \beta^{a}_{\mu}\partial_{\nu} \omega_{\lambda}, 
\end{align}   
with the same $K$-matrix in the first term of ${\mathcal L}_{\rm topo}$ appearing in flux attachment Eq.\eqref{CB:CS}. The Hall viscosity and Wen-Zee term are reproduced correctly here.\cite{Hansson2013, Gromov2014b} It is straightforward to generalize to the other abelian multi-component FQH states. 

\section{Projective Parton Constructions} 
\label{sec:parton}

We will now discuss the Hall viscosity and geometric responses of abelian and non-abelian FQH states using the projective parton construction of Refs. [\onlinecite{Wen1999}] and [\onlinecite{ Barkeshli2010}]. In this picture the electron is formally split into several ``partons'', each with a certain preassigned charge and all coupled to the same uniform magnetic field. This formal enlargement of the Hilbert space leads to a new local gauge symmetry. The action of the associated gauge fields projects the Hilbert space into the physical subspace of the original fermions. This procedure yields a correct effective theory in all cases\cite{Wen1999,Barkeshli2010} but, as we will see below, has some open issues in the non-abelian case.

\subsection{Abelian States}

We begin with the projective parton description of the Laughlin state $\nu = \frac{1}{2p+1}, p \in {\mathbb Z}$. In this construction the  electron operator factorizes into $2p+1$ fermionic partons\cite{Wen1999,Barkeshli2012}
\beq
\Psi_{e} (z) = \psi_{1} (z) \psi_{2} (z) \cdots \psi_{2p+1}(z)
\label{electron:abelian}
\eeq  
which is a singlet under a local $SU(2p+1)$ gauge symmetry. These ``emergent'' gauge symmetries are characteristic of parton constructions.
Each parton $\psi_{i}$ carries the fractional electric charge $e/(2p+1)$ and fills up a lowest Landau level. Notice that the electron and the partons are all scalars and thus do not couple with the spin connection minimally. We also need to introduce $2p$ internal $U(1)$ gauge fields (or a SU$(2p+1)$ gauge field) to project out the non-physical states in the Hilbert space spanned by the partons of Eq.\eqref{electron:abelian}.\cite{Wen1999} We assume that the partons see the same background metric as the electron. As each parton is in the $\nu=1$ state and is gapped, we integrate out the partons to 
express the result in terms of a hydrodynamic theory of the Laughlin state. The resulting theory is identical to those of the composite particle theories, {\it e.g.} Eq.\eqref{sec2sub2:hydrodynamic}, except the gravitational CS term. Hence, we find that the correct Hall viscosity and Wen-Zee term are reproduced in the projective parton approach, but the central charge is overestimated as $c= 2p+1$. 

As a concrete example of this, we study the bosonic Laughlin state at $\nu = \frac{1}{2}$. For this state, we fractionalize a bosonic field $b$ into the two fermionic partons $\psi_{i}, i =1,2$ carrying $\frac{1}{2}$ electric charge. 
\beq
b(z) = \psi_{1}(z)\psi_{2}(z) 
\label{boson}
\eeq
The Hilbert space of the partons $\psi_{i}$ has unphysical states, and we need to project out those unphysical states by requiring that $\rho_{b} = \langle b^{\dagger} b \rangle$ and $\rho^{\psi}_{j} = \langle \psi^{\dagger}_{j}\psi_{j} \rangle , j =1, 2$ are the same, {\it i.e.}  $\rho_{b} = \rho^{\psi}_{j}, j=1,2$. This projection can be implemented by introducing an internal $U(1)$ gauge field $a_{\mu}$. Under the $U(1)$ gauge field,~\cite{Wen1999,Barkeshli2012}  $\psi_{1}$ and $\psi_{2}$ are oppositely charged because the fundamental boson $b$ should be invariant under the $U(1)$ gauge transformation. To describe the Laughlin state, we choose the mean field state where the fermionic partons $\psi_{i}$ are in $\nu = 1$ state. Furthermore, the partons are scalars and thus do not minimally couple with the spin connection. 
\begin{align}
{\mathcal L} &= \sum_{j=1}^{2}\sqrt{g} \bigg[ \frac{i}{2}\left( \left( D^{j}_{0} \psi_{j}(x) \right)^{\dagger} \Psi_{j} - \Psi^{\dagger}_{j}(x) \left( D^{j}_{0} \Psi_{j}(x) \right)  \right) \nonumber\\
& - \frac{1}{2} (D^{j}_{a}\Psi_{j} (x))^{\dagger}g^{ab}(D^{j}_{b}\Psi_{j} (x) ) \bigg] 
\label{boson:action}
\end{align}
 in which $D^{j}_{\mu} = \partial_{\mu} + i\frac{1}{2}A_{\mu}\pm i a_{\mu}$ are the covariant derivatives of the fermionic partons $\psi_{j}, j =1,2$ ($+ i a_{\mu}$ for $\psi_{1}$ and $-ia_{\mu}$ for $\psi_{2}$). We integrate out the partons and obtain the effective theory 
\begin{align}
{\mathcal L} =&  \rho_{b} \omega_{t} + \rho_{b} A_{t} + \frac{2}{4\pi} \varepsilon^{\mu\nu\lambda} a_{\mu}\partial_{\nu} a_{\lambda}  \nonumber\\ 
&+ \frac{1}{2}\frac{1}{4\pi} \varepsilon^{\mu\nu\lambda} \bigg( A_{\mu} + {\bar s} \omega_{\mu} \bigg) \partial_{\nu} \bigg( A_{\lambda} + {\bar s} \omega_{\lambda} \bigg)  + \cdots, \nonumber\\ 
\label{boson:effective}
\end{align}
with the average orbital spin ${\bar s} = 1$. The effective theory is obtained by replacing $\delta A_{\mu}$ in Eq.\eqref{sec2sub2:integer} (the effective theory of the integer quantum Hall fluid) by $\frac{1}{2}A_{\mu}\pm a_{\mu}$ because each parton is at the filling $\nu = 1$ and minimally couples to $\frac{1}{2}A_{\mu}\pm a_{\mu}$. The average orbital spin can be deduced from the coefficients of the mutual CS term between $A_{\mu}$ and $\omega_{\mu}$. We also find the correct Hall viscosity from the effective action. 
\beq
\eta_{H} = \frac{\rho_{b}}{2} 
\label{abelian:final}
\eeq
This is consistent with the average orbital spin $s = 1$. 

In fact, it is better to recast Eq. \eqref{boson:effective} into the following form which is more amenable to be turned into the hydrodynamic description 
\begin{align}
&{\mathcal L} =  \rho_{b} \omega_{t} + \rho_{b} A_{t} + \frac{\varepsilon^{\mu\nu\lambda}}{4\pi} \alpha_{\mu}\partial_{\nu}\alpha_{\lambda} + \frac{\varepsilon^{\mu\nu\lambda}}{4\pi}\beta_{\mu}\partial_{\nu}\beta_{\lambda}  \nonumber\\ 
&~ - \frac{2 }{48\pi} \varepsilon^{\mu\nu\lambda} \omega_{\mu}\partial_{\nu}\omega_{\lambda} + \cdots, 
\label{boson:effective2}
\end{align}
in which $ \alpha_{\mu} = \frac{1}{2}A_{\mu} + \frac{1}{2}\omega_{\mu} + a_{\mu}$ and $\beta_{\mu} = \frac{1}{2}A_{\mu} + \frac{1}{2}\omega_{\mu} - a_{\mu}$. Then we introduce the {\it two} hydrodynamic fields $b_{1, \mu}$ and $b_{2, \mu}$ to rewrite the CS terms of $\alpha_{\mu}$ and $\beta_{\mu}$ appearing in Eq.\eqref{boson:effective2}
\begin{align}
 &{\mathcal L} =  \rho_{b} \omega_{t} + \rho_{b} A_{t} - \frac{\varepsilon^{\mu\nu\lambda}}{4\pi} b_{1,\mu}\partial_{\nu}b_{1,\lambda} - \frac{\varepsilon^{\mu\nu\lambda}}{4\pi}b_{2,\mu}\partial_{\nu}b_{2,\lambda}\nonumber\\ 
& +\frac{\varepsilon^{\mu\nu\lambda}}{2\pi} b_{1,\mu} \partial_{\nu} \alpha_{\lambda} + \frac{\varepsilon^{\mu\nu\lambda}}{2\pi} b_{2,\mu} \partial_{\nu} \beta_{\lambda} - \frac{2 }{48\pi} \varepsilon^{\mu\nu\lambda} \omega_{\mu}\partial_{\nu}\omega_{\lambda} 
\label{boson:effective3}
\end{align}
We integrate out $a_{\mu}$ and obtain the equation of motion $b_{1,\mu} = b_{2,\mu}$. We set $b_{\mu} = b_{1,\mu}= b_{2,\mu}$ and then \eqref{boson:effective3} becomes  
\begin{align}
 &{\mathcal L} =  \rho_{b} \omega_{t} + \rho_{b} A_{t} - \frac{2}{4\pi}\varepsilon^{\mu\nu\lambda} b_{\mu}\partial_{\nu}b_{\lambda} +\frac{\varepsilon^{\mu\nu\lambda}}{2\pi} b_{\mu} \partial_{\nu} A_{\lambda}  \nonumber\\ 
&+ \frac{2}{4\pi}\varepsilon^{\mu\nu\lambda} b_{\mu} \partial_{\nu} \omega_{\lambda} - \frac{2 }{48\pi} \varepsilon^{\mu\nu\lambda} \omega_{\mu}\partial_{\nu}\omega_{\lambda},
\end{align}
where we notice that the Hall viscosity and Wen-Zee term are correctly captured in this projective parton approach. However, it yields the wrong central charge $2$, which is doubly larger than the correct value. 

Now, for the general Laughlin state at $\nu = \frac{1}{k}, k \in {\mathbb Z}$, we start with the definition for the fundamental particle with $k$ partons carrying the electromagnetic charge $\frac{1}{k}$. 
\beq
\Psi_{e} (z) = \psi_{1} (z) \psi_{2} (z) \cdots \psi_{k}(z)
\eeq 
To describe the Laughlin state, each parton should be at the filling $\nu =1$. This mean field ansatz and the fundamental particle are invariant under the $(k-1)$ internal $U(1)$ gauge fields $(a_{1,\mu}, a_{2,\mu} \cdots, a_{k-1, \mu} )$. We choose the coupling between the gauge fields and the partons in the way that the $j$-th parton $\psi_{j}$ ($1<j<k$) couples minimally to $\alpha_{j, \mu} = a_{j-1,\mu}- a_{j,\mu}$, and the 1st parton $\psi_{1}$ (the last parton $\psi_{k}$) couples only to $\alpha_{1,\mu} = -a_{1,\mu}$ ($\alpha_{k-1, \mu} = a_{k-1,\mu}$). It is convenient to introduce another set of the gauge fields $(\beta_{1, \mu}, \beta_{2, \mu}, \cdots, \beta_{k,\mu})$ such that 
\beq
\beta_{j, \mu} = \alpha_{j,\mu} + \frac{1}{k} A_{\mu} + \frac{1}{2}\omega_{\mu}. 
\eeq
We integrate out the partons to obtain the effective response theory, 
\begin{align}
&{\mathcal L} = \sum_{i=1}^{k} \bigg(\rho_{i}\frac{1}{k}A_{t} + \frac{1}{2} \rho_{i} \omega_{t}\bigg) + \sum_{i=1}^{k} \frac{\varepsilon^{\mu\nu\lambda}}{4\pi} \beta_{i, \mu}\partial_{\nu} \beta_{i, \lambda}.
\end{align}
Then we introduce the hydrodynamic fields $b_{i, \mu}$ to rewrite the CS terms. 
\begin{align}
&{\mathcal L} = \sum_{i=1}^{k} \bigg(\rho_{i}\frac{1}{k}A_{t} + \frac{1}{2} \rho_{i} \omega_{t}\bigg) - \sum_{i=1}^{k}\frac{\varepsilon^{\mu\nu\lambda}}{4\pi} b_{i, \mu}\partial_{\nu} b_{i, \lambda} \nonumber\\
&~ +\sum_{i=1}^{k}\frac{\varepsilon^{\mu\nu\lambda}}{2\pi} b_{i, \mu}\partial_{\nu} \beta_{i, \lambda}
\end{align}
Now we integrate out $a_{j, \mu}, j = 1, 2 \cdots k-1$ and this generates the equation of motion $b_{\mu} = b_{i, \mu}, i = 1, 2, \cdots k$. Then we find the same effective hydrodynamic response theory as that of the composite fermion Eq.\eqref{sec2sub2:hydrodynamic} in the main text except the overestimate of the chiral central charge by $k$ times, {\it i.e.,} we will find a wrong central charge $c=k$ for the Laughlin state instead of the correct value $c=1$. Notice that $k$ is the central charge of the mean-field state. Hence, the projective parton construction predicts a wrong central charge, which is of the mean-field state. %We suspect that this overestimate of the central charge seems to suggest that, within the approximation prescribed in Ref.[\onlinecite{Wen1999}], the partons are not ``projected" properly and that each parton behaves as an independent dynamical degree of freedom and contribute to the energy-momentum current, contrast to the expectation.\cite{Wen1999} However, at this point, it is not clear where the overestimate of the chiral central charge comes from and how to correct this overestimate. 

\subsection{Non-abelian States} 
\label{sec:non-abelian}

Wen \cite{Wen1999} (and Barkeshli and Wen \cite{Barkeshli2010})  generalized the parton construction for the non-abelian ${\mathbb Z}_{k}$ Read-Rezayi parafermion states \cite{Read1999} (including the $k=2$ fermionic and bosonic pfaffian states) at  filling $\nu = \frac{k}{Mk+2}$. The fundamental particle $\Psi_{e}$ now is
% becomes
\beq
\Psi_{e} (z)= \psi_{1}\psi_{2} \cdots \psi_{M} \times \sum^{k}_{a=1} f_{2a-1}f_{2a}.
\label{electron:nonabelian}
\eeq
Here $\psi_{i}, i =1,2 \cdots M$ carries electric charge $q_{\psi} = \frac{k}{Mk+2}$, and $f_{a}, a= 1 \cdots 2k$ carries electric charge $q_{f}=\frac{1}{Mk+2}$. Thus we introduce the electromagnetic charge matrix, 
\begin{align}
Q= 
\begin{pmatrix} 
q_{\psi} I_{M \times M} & 0  \\
0 & q_{f} I_{2k \times 2k}
\end{pmatrix}.
\end{align}
All the partons $\psi_{i}$ and $f_{j}$ are fermions in a $\nu =1$ state. The state has $U(M) \times Sp(2k)$ gauge symmetry, under which the electron operator of Eq.\eqref{electron:nonabelian} is invariant. This construction of the electron operator satisfies $SU(2)_k$ current algebra \cite{Barkeshli2010}  and  generates the same ${\mathbb Z}_{k}$ parafermion state wavefunction as the $SU(2)_k$ Wess-Zumino-Witten conformal field theory. By integrating out the partons, we obtain the effective field theory
\begin{align}
{\mathcal L} &=  {\text Tr} \bigg( (Q \delta A_0 + a_0) \rho \bigg) +\frac{1}{2} {\text Tr}(\rho) \omega_0 \nonumber\\ 
&+ \frac{1}{4\pi} {\text Tr}(Q^2) \varepsilon^{\mu\nu\lambda}\delta A_{\mu}\partial_{\nu} \delta A_{\lambda} + \frac{\varepsilon^{\mu\nu\lambda}}{4\pi} \delta A_{\mu}{\text Tr}(Q {\mathcal F}_{\nu\lambda} ) \nonumber\\
&+\frac{\varepsilon^{\mu\nu\lambda}}{8\pi} {\text Tr}(a_{\mu} {\mathcal F}_{\nu\lambda}) +  \frac{\varepsilon^{\mu\nu\lambda}}{8\pi} \omega_{\mu}{\text Tr}(Q F_{\nu\lambda} + {\mathcal F}_{\nu\lambda})
\label{effective:nonabelian}
\end{align}
where
\begin{equation}
 \rho = \begin{pmatrix} 
\rho_{\psi} I_{M \times M} & 0  \\
0 & \rho_{f} I_{2k \times 2k}
\end{pmatrix},
\end{equation}
$F_{\nu\lambda} = \partial_{\nu} \delta A_{\lambda} - \partial_{\lambda}\delta A_{\nu}$, ${\mathcal F}_{\nu\lambda}$ is the field strength of $a_{\mu} \in U(M) \times Sp(2k)$, and $\rho^{\psi} = \rho_{e}$ and $\rho^{f} = \rho_{e}/k$. Using Tr$({\mathcal F}_{\nu\lambda}) = 0$,  we find the effective response %theory 
%${\mathcal L}_{\rm eff}$ 
of the FQH state to the external electromagnetic gauge field $\delta A_{\mu}$ and a distortion of the geometry
\begin{align}
{\mathcal L}_{\rm eff} &=  \rho_{e}\delta A_{0} + \frac{M+2}{2}\rho_{e} \omega_{0}\nonumber\\ 
&+ \frac{k}{Mk+2} \frac{1}{4\pi} \varepsilon^{\mu\nu\lambda}(\delta A_{\mu}+ {\bar s} \omega_{\mu})\partial_{\nu} (\delta A_{\lambda}+{\bar s} \omega_{\lambda})  + \cdots
\label{nonabelian-final}
\end{align}
where ${\bar s} = (M+2)/2$.
This effective Lagrangian 
%correctly 
yields the average orbital spin ${\bar s}$ for the non-abelian FQH state, and  the Berry phase term, $\frac{M+2}{2} \rho \omega_{0}$, in Eq.\eqref{nonabelian-final}, yields the correct Hall viscosity.\cite{Read2011} However, the  gravitational CS term of Eq.\eqref{nonabelian-final}  is an integer %instead of a  fractional value which is 
although it
% is expected to
should be fractional 
%from the theory of the theory of the edge states of 
for non-abelian FQH fluids. %The preceding

The parton approach can be generalized to general FQH states which have the parton description. The preceding parton approach can be generalized to compute  the effective response of  general FQH states.\cite{Wen1999} Let us consider a set of fermionic partons $\{ f_{1}, f_{2} \cdots f_{K} \}$ with a definition for the electron operator {\it e.g.,} that of Eq.\eqref{electron:nonabelian}. Each parton $f_{i}$ carries the electric charge $q_{i}$ such that the $K \times K$ electromagnetic charge matrix is given by $Q = q_i \delta_{ij}$. If not further structure is assumed, the partons are all scalars and thus do not couple minimally to the spin connection. The partons may have  different integer filling $m_{i} \in {\mathbb Z}$, and hence we also define a $K \times K$ filling matrix $M = m_{i} \delta_{ij}$. This sets the spin {\it matrix} for partons as $S = \frac{m_{i}}{2} \delta_{ij}$. The density of partons is the matrix ${\rho} = \rho_{i}\delta_{ij}$. Furthermore we assume that the partons see the same background metric as the electrons. 
From these assumptions, we deduce that there is a gauge group $G$ which leaves the electron operator and this mean-field state invariant. Thus the internal or statistical gauge field $a_{\mu}$ lives in the algebra of $G$. As the partons are in  integer quantum Hall states, they are gapped and can be integrated out to find an effective theory of the same form as Eq.\eqref{effective:nonabelian} except that the partons couple with a spin matrix $S$.
\begin{align}
{\mathcal L} =&  {\text Tr} \bigg( \rho(Q A_{t} + a_{t})\bigg) + {\text Tr}(\rho S) \omega_{t} \nonumber\\
+& \frac{1}{4\pi} {\text Tr}(Q^2 M) \varepsilon^{\mu\nu\lambda}A_{\mu}\partial_{\nu}A_{\lambda} + \frac{\varepsilon^{\mu\nu\lambda}}{4\pi} A_{\mu}{\text Tr}(MQ {\mathcal F}_{\nu\lambda} ) \nonumber\\
+&\frac{\varepsilon^{\mu\nu\lambda}}{8\pi} {\text Tr}(M a_{\mu} {\mathcal F}_{\nu\lambda}) +  \frac{\varepsilon^{\mu\nu\lambda}}{4\pi} \omega_{\mu}{\text Tr}\bigg(MS (Q F_{\nu\lambda} + {\mathcal F}_{\nu\lambda})\bigg) 
\end{align}
in which $F_{\nu\lambda} = \partial_{\nu} A_{\lambda} - \partial_{\lambda} A_{\nu}$ and ${\mathcal F}_{\nu\lambda}$ is the field strength of $a_{\mu}$. To obtain this effective action, we first assumed that the gauge field $a_{\mu}$ is taken from the maximally abelian subgroup of the gauge group and integrate out the fermionic partons. Because the mean-field state does not break the gauge symmetry, we can restore the full gauge invariant action.\cite{Wen1999} Here too, the projective parton approach does not yield  the consistent value of the gravitational CS term.\cite{Gromov2014, Gromov2014b}

\section{Conclusions}
\label{sec:conclusions}

%In conclusion, 
We  derived a theory of the Hall viscosity and Wen-Zee term for FQH states from the composite particle theories and the projective parton approach.  
%We  showed that the 
The composite particles carry the spin because of the spin-statistics connection, and %they 
couple with the background geometry through the spin connection. %With this knowledge, we 
We derived 
% successfully  
the correct Hall viscosity and Wen-Zee term
%geometric response %within the 
for CF and CB theories.
%, {\it i.e.} the  Hall viscosity and Wen-Zee term. 
In the projective parton construction, we 
%showed how to obtain 
obtained the electromagnetic and geometric response of general FQH states, both abelian and  non-abelian. %Because 
%Since the geometric response depends only on the orbital spin, which is required to identify the topological order of FQH states, 
%our results imply that the 
%the field theory yields the consistent descriptions of topological order and  the geometric response.  
%We also showed that  the 
%The CF approach yields the correct gravitational CS term (for abelian FQH states), but the
We found that the composite particle theories and the projective parton approach do not yield the correct gravitational CS term,
% This is to be contrasted with 
while the universal global ground state properties, such as the Hall conductivity and the ground state degeneracy, 
%which 
are correctly reproduced in all cases. 
% The difference between the CF and the CB approaches reflects the fact that   the mean-field CF theory breaks time-reversal invariance %at the mean-field level (although 
%(with the wrong current algebra) while the  mean-field CB theory does not.% (at the mean field level). 
%Instead, the 
%The parton construction yields a value of the central charge which is inconsistent with the CS theory of the hydrodynamic field and with theory of the edge states. 
%Aside from this  this issue, in this work we provided a firm ground to study the geometric responses of FQH states, and we hope that our work sheds light on the studies on the geometric response, and the relation between  2+1-dimensional gravity  and FQH states.

\begin{acknowledgments}
We  thank T. Hughes, D. T. Son and J. Teo  for helpful discussions. This work is supported in part by the by the NSF grants numbers  DMR-1064319 and DMR-1408713 at UIUC and by ICMT.
\end{acknowledgments}

%\bibliographystyle{apsrev}
%\bibliography{Geometric.bib}

\begin{thebibliography}{53}
\expandafter\ifx\csname natexlab\endcsname\relax\def\natexlab#1{#1}\fi
\expandafter\ifx\csname bibnamefont\endcsname\relax
  \def\bibnamefont#1{#1}\fi
\expandafter\ifx\csname bibfnamefont\endcsname\relax
  \def\bibfnamefont#1{#1}\fi
\expandafter\ifx\csname citenamefont\endcsname\relax
  \def\citenamefont#1{#1}\fi
\expandafter\ifx\csname url\endcsname\relax
  \def\url#1{\texttt{#1}}\fi
\expandafter\ifx\csname urlprefix\endcsname\relax\def\urlprefix{URL }\fi
\providecommand{\bibinfo}[2]{#2}
\providecommand{\eprint}[2][]{\url{#2}}

\bibitem[{\citenamefont{{Laughlin}}(1983)}]{Laughlin83}
\bibinfo{author}{\bibfnamefont{R.~B.} \bibnamefont{{Laughlin}}},
  \bibinfo{journal}{Phys. Rev. Lett.} \textbf{\bibinfo{volume}{50}},
  \bibinfo{pages}{1395} (\bibinfo{year}{1983}).

\bibitem[{\citenamefont{Zhang}(1992)}]{Zhang1992}
\bibinfo{author}{\bibfnamefont{S.~C.} \bibnamefont{Zhang}},
  \bibinfo{journal}{Int. J. Mod. Phys. B} \textbf{\bibinfo{volume}{6}},
  \bibinfo{pages}{25} (\bibinfo{year}{1992}).

\bibitem[{\citenamefont{Haldane}(1983)}]{Haldane1983}
\bibinfo{author}{\bibfnamefont{F.~D.~M.} \bibnamefont{Haldane}},
  \bibinfo{journal}{Phys. Rev. Lett.} \textbf{\bibinfo{volume}{51}},
  \bibinfo{pages}{605} (\bibinfo{year}{1983}).

\bibitem[{\citenamefont{Jain}(1989)}]{Jain1989}
\bibinfo{author}{\bibfnamefont{J.~K.} \bibnamefont{Jain}},
  \bibinfo{journal}{Phys. Rev. Lett.} \textbf{\bibinfo{volume}{63}},
  \bibinfo{pages}{199} (\bibinfo{year}{1989}).

\bibitem[{\citenamefont{L{\'o}pez and Fradkin}(1991)}]{Lopez1991}
\bibinfo{author}{\bibfnamefont{A.}~\bibnamefont{L{\'o}pez}} \bibnamefont{and}
  \bibinfo{author}{\bibfnamefont{E.}~\bibnamefont{Fradkin}},
  \bibinfo{journal}{Phys. Rev. B} \textbf{\bibinfo{volume}{44}},
  \bibinfo{pages}{5246} (\bibinfo{year}{1991}).

\bibitem[{\citenamefont{Wen}(1995)}]{Wen95}
\bibinfo{author}{\bibfnamefont{X.~G.} \bibnamefont{Wen}},
  \bibinfo{journal}{Adv. Phys.} \textbf{\bibinfo{volume}{44}},
  \bibinfo{pages}{405} (\bibinfo{year}{1995}).

\bibitem[{\citenamefont{Wen}(1992)}]{wenrev1}
\bibinfo{author}{\bibfnamefont{X.-G.} \bibnamefont{Wen}},
  \bibinfo{journal}{Int. J. Mod. Phys. B} \textbf{\bibinfo{volume}{6}},
  \bibinfo{pages}{1711} (\bibinfo{year}{1992}).

\bibitem[{\citenamefont{Wen}(2004)}]{Wen2004book}
\bibinfo{author}{\bibfnamefont{X.-G.} \bibnamefont{Wen}},
  \emph{\bibinfo{title}{Quantum field theory of many-body systems: from the
  origin of sound to an origin of light and electrons}}
  (\bibinfo{publisher}{Oxford University Press Oxford}, \bibinfo{year}{2004}).

\bibitem[{\citenamefont{Fradkin}(2013)}]{fradkin-1991}
\bibinfo{author}{\bibfnamefont{E.}~\bibnamefont{Fradkin}},
  \emph{\bibinfo{title}{Field Theories of Condensed Matter Systems}}
  (\bibinfo{publisher}{Cambridge University Press},
  \bibinfo{address}{Cambridge, UK}, \bibinfo{year}{2013}),
  \bibinfo{edition}{2nd} ed.

\bibitem[{\citenamefont{Wen}(2012)}]{Wen2012}
\bibinfo{author}{\bibfnamefont{X.-G.} \bibnamefont{Wen}},
  \emph{\bibinfo{title}{Modular transformation and bosonic/fermionic
  topological orders in {Abelian} fractional quantum {Hall} states}}
  (\bibinfo{year}{2012}), \bibinfo{note}{unpublished},
  \eprint{arXiv:1212.5121}.

\bibitem[{\citenamefont{Wen and Zee}(1992{\natexlab{a}})}]{wenzee}
\bibinfo{author}{\bibfnamefont{X.~G.} \bibnamefont{Wen}} \bibnamefont{and}
  \bibinfo{author}{\bibfnamefont{A.}~\bibnamefont{Zee}},
  \bibinfo{journal}{Phys. Rev. Lett.} \textbf{\bibinfo{volume}{69}},
  \bibinfo{pages}{953} (\bibinfo{year}{1992}{\natexlab{a}}).

\bibitem[{\citenamefont{Read}(2009)}]{Read2009}
\bibinfo{author}{\bibfnamefont{N.}~\bibnamefont{Read}}, \bibinfo{journal}{Phys.
  Rev. B} \textbf{\bibinfo{volume}{79}}, \bibinfo{pages}{045308}
  (\bibinfo{year}{2009}).

\bibitem[{\citenamefont{Read and Rezayi}(2011)}]{Read2011}
\bibinfo{author}{\bibfnamefont{N.}~\bibnamefont{Read}} \bibnamefont{and}
  \bibinfo{author}{\bibfnamefont{E.~H.} \bibnamefont{Rezayi}},
  \bibinfo{journal}{Phys. Rev. B} \textbf{\bibinfo{volume}{84}},
  \bibinfo{pages}{085316} (\bibinfo{year}{2011}).

\bibitem[{\citenamefont{Hoyos and Son}(2012)}]{Hoyos2012}
\bibinfo{author}{\bibfnamefont{C.}~\bibnamefont{Hoyos}} \bibnamefont{and}
  \bibinfo{author}{\bibfnamefont{D.~T.} \bibnamefont{Son}},
  \bibinfo{journal}{Phys. Rev. Lett.} \textbf{\bibinfo{volume}{108}},
  \bibinfo{pages}{066805} (\bibinfo{year}{2012}).

\bibitem[{\citenamefont{Bradlyn et~al.}(2012)\citenamefont{Bradlyn, Goldstein,
  and Read}}]{Bradlyn2012}
\bibinfo{author}{\bibfnamefont{B.}~\bibnamefont{Bradlyn}},
  \bibinfo{author}{\bibfnamefont{M.}~\bibnamefont{Goldstein}},
  \bibnamefont{and} \bibinfo{author}{\bibfnamefont{N.}~\bibnamefont{Read}},
  \bibinfo{journal}{Phys. Rev. B} \textbf{\bibinfo{volume}{86}},
  \bibinfo{pages}{245309} (\bibinfo{year}{2012}).

\bibitem[{\citenamefont{Haldane}(2009)}]{Haldane2009}
\bibinfo{author}{\bibfnamefont{F.~D.~M.} \bibnamefont{Haldane}},
  \emph{\bibinfo{title}{{``Hall viscosity'' and intrinsic metric of
  incompressible fractional Hall fluids}}} (\bibinfo{year}{2009}),
  \bibinfo{note}{unpublished}, \eprint{arXiv:0906.1854}.

\bibitem[{\citenamefont{Hughes et~al.}(2011)\citenamefont{Hughes, Leigh, and
  Fradkin}}]{Hughes2011}
\bibinfo{author}{\bibfnamefont{T.~L.} \bibnamefont{Hughes}},
  \bibinfo{author}{\bibfnamefont{R.~G.} \bibnamefont{Leigh}}, \bibnamefont{and}
  \bibinfo{author}{\bibfnamefont{E.}~\bibnamefont{Fradkin}},
  \bibinfo{journal}{Phys. Rev. Lett.} \textbf{\bibinfo{volume}{107}},
  \bibinfo{pages}{075502} (\bibinfo{year}{2011}).

\bibitem[{\citenamefont{Hughes et~al.}(2013)\citenamefont{Hughes, Leigh, and
  Parrikar}}]{Hughes2013}
\bibinfo{author}{\bibfnamefont{T.~L.} \bibnamefont{Hughes}},
  \bibinfo{author}{\bibfnamefont{R.~G.} \bibnamefont{Leigh}}, \bibnamefont{and}
  \bibinfo{author}{\bibfnamefont{O.}~\bibnamefont{Parrikar}},
  \bibinfo{journal}{Phys. Rev. D} \textbf{\bibinfo{volume}{88}},
  \bibinfo{pages}{025040} (\bibinfo{year}{2013}).

\bibitem[{\citenamefont{Abanov and Gromov}(2014)}]{Gromov2014}
\bibinfo{author}{\bibfnamefont{A.~G.} \bibnamefont{Abanov}} \bibnamefont{and}
  \bibinfo{author}{\bibfnamefont{A.}~\bibnamefont{Gromov}},
  \bibinfo{journal}{Phys. Rev. B} \textbf{\bibinfo{volume}{90}},
  \bibinfo{pages}{014435} (\bibinfo{year}{2014}).

\bibitem[{\citenamefont{Gromov and Abanov}(2014)}]{Gromov2014b}
\bibinfo{author}{\bibfnamefont{A.}~\bibnamefont{Gromov}} \bibnamefont{and}
  \bibinfo{author}{\bibfnamefont{A.~G.} \bibnamefont{Abanov}},
  \emph{\bibinfo{title}{{Density-curvature response and gravitational
  anomaly}}} (\bibinfo{year}{2014}), \bibinfo{note}{unpublished},
  \eprint{arXiv:1403.5809}.

\bibitem[{\citenamefont{Nicolis and Son}(2011)}]{Nicolis2011}
\bibinfo{author}{\bibfnamefont{A.}~\bibnamefont{Nicolis}} \bibnamefont{and}
  \bibinfo{author}{\bibfnamefont{D.~T.} \bibnamefont{Son}},
  \emph{\bibinfo{title}{{Hall viscosity from effective field theory}}}
  (\bibinfo{year}{2011}), \bibinfo{note}{unpublished},
  \eprint{arXiv:1103.2137}.

\bibitem[{\citenamefont{Son}(2013)}]{Son2013}
\bibinfo{author}{\bibfnamefont{D.~T.} \bibnamefont{Son}},
  \emph{\bibinfo{title}{{Newton-Cartan Geometry and the Quantum Hall Effect}}}
  (\bibinfo{year}{2013}), \bibinfo{note}{unpublished},
  \eprint{arXiv:1306.0638}.

\bibitem[{\citenamefont{Avron et~al.}(1995)\citenamefont{Avron, Seiler, and
  Zograf}}]{Avron1995}
\bibinfo{author}{\bibfnamefont{J.~E.} \bibnamefont{Avron}},
  \bibinfo{author}{\bibfnamefont{R.}~\bibnamefont{Seiler}}, \bibnamefont{and}
  \bibinfo{author}{\bibfnamefont{P.~G.} \bibnamefont{Zograf}},
  \bibinfo{journal}{Phys. Rev. Lett.} \textbf{\bibinfo{volume}{75}},
  \bibinfo{pages}{697} (\bibinfo{year}{1995}).

\bibitem[{\citenamefont{Fremling et~al.}(2013)\citenamefont{Fremling, Hansson,
  and Suorsa}}]{Hansson2013}
\bibinfo{author}{\bibfnamefont{M.}~\bibnamefont{Fremling}},
  \bibinfo{author}{\bibfnamefont{T.~H.} \bibnamefont{Hansson}},
  \bibnamefont{and} \bibinfo{author}{\bibfnamefont{J.}~\bibnamefont{Suorsa}},
  \bibinfo{journal}{Phys. Rev. B} \textbf{\bibinfo{volume}{89}},
  \bibinfo{pages}{125303} (\bibinfo{year}{2013}).

\bibitem[{\citenamefont{Wen and Zee}(1992{\natexlab{b}})}]{Wen1992}
\bibinfo{author}{\bibfnamefont{X.~G.} \bibnamefont{Wen}} \bibnamefont{and}
  \bibinfo{author}{\bibfnamefont{A.}~\bibnamefont{Zee}},
  \bibinfo{journal}{Phys. Rev. B} \textbf{\bibinfo{volume}{46}},
  \bibinfo{pages}{2290} (\bibinfo{year}{1992}{\natexlab{b}}).

\bibitem[{\citenamefont{Halperin}(1984)}]{Halperin1984}
\bibinfo{author}{\bibfnamefont{B.~I.} \bibnamefont{Halperin}},
  \bibinfo{journal}{Phys. Rev. Lett.} \textbf{\bibinfo{volume}{52}},
  \bibinfo{pages}{1583} (\bibinfo{year}{1984}).

\bibitem[{\citenamefont{Moore and Read}(1991)}]{Moore1991}
\bibinfo{author}{\bibfnamefont{G.}~\bibnamefont{Moore}} \bibnamefont{and}
  \bibinfo{author}{\bibfnamefont{N.}~\bibnamefont{Read}},
  \bibinfo{journal}{Nucl. Phys. B} \textbf{\bibinfo{volume}{360}},
  \bibinfo{pages}{362} (\bibinfo{year}{1991}).

\bibitem[{\citenamefont{Read and Rezayi}(1999)}]{Read1999}
\bibinfo{author}{\bibfnamefont{N.}~\bibnamefont{Read}} \bibnamefont{and}
  \bibinfo{author}{\bibfnamefont{E.}~\bibnamefont{Rezayi}},
  \bibinfo{journal}{Phys. Rev. B} \textbf{\bibinfo{volume}{59}},
  \bibinfo{pages}{8084} (\bibinfo{year}{1999}).

\bibitem[{\citenamefont{Zhang et~al.}(1989)\citenamefont{Zhang, Hansson, and
  Kivelson}}]{Zhang1989}
\bibinfo{author}{\bibfnamefont{S.~C.} \bibnamefont{Zhang}},
  \bibinfo{author}{\bibfnamefont{T.~H.} \bibnamefont{Hansson}},
  \bibnamefont{and} \bibinfo{author}{\bibfnamefont{S.}~\bibnamefont{Kivelson}},
  \bibinfo{journal}{Phys. Rev. Lett.} \textbf{\bibinfo{volume}{62}},
  \bibinfo{pages}{82} (\bibinfo{year}{1989}).

\bibitem[{\citenamefont{Wen}(1999)}]{Wen1999}
\bibinfo{author}{\bibfnamefont{X.-G.} \bibnamefont{Wen}},
  \bibinfo{journal}{Phys. Rev. B} \textbf{\bibinfo{volume}{60}},
  \bibinfo{pages}{8827} (\bibinfo{year}{1999}).

\bibitem[{\citenamefont{Saremi and Son}(2012)}]{saremi-2012}
\bibinfo{author}{\bibfnamefont{O.}~\bibnamefont{Saremi}} \bibnamefont{and}
  \bibinfo{author}{\bibfnamefont{D.~T.} \bibnamefont{Son}},
  \bibinfo{journal}{Journal of High Energy Physics}
  \textbf{\bibinfo{volume}{2012}}, \bibinfo{pages}{1} (\bibinfo{year}{2012}).

\bibitem[{\citenamefont{Son and Wu}(2013)}]{son-2013-holographic}
\bibinfo{author}{\bibfnamefont{D.~T.} \bibnamefont{Son}} \bibnamefont{and}
  \bibinfo{author}{\bibfnamefont{C.}~\bibnamefont{Wu}},
  \emph{\bibinfo{title}{{Holographic Spontaneous Parity Breaking and Emergent
  Hall Viscosity and Angular Momentum}}} (\bibinfo{year}{2013}),
  \bibinfo{note}{unpublished}, \eprint{arXiv:1311.4882}.

\bibitem[{\citenamefont{Chen et~al.}(2012{\natexlab{a}})\citenamefont{Chen,
  Dai, Lee, and Maity}}]{chen-2012a}
\bibinfo{author}{\bibfnamefont{J.-W.} \bibnamefont{Chen}},
  \bibinfo{author}{\bibfnamefont{S.-H.} \bibnamefont{Dai}},
  \bibinfo{author}{\bibfnamefont{N.-E.} \bibnamefont{Lee}}, \bibnamefont{and}
  \bibinfo{author}{\bibfnamefont{D.}~\bibnamefont{Maity}},
  \bibinfo{journal}{Journal of High Energy Physics}
  \textbf{\bibinfo{volume}{2012}}, \bibinfo{pages}{1}
  (\bibinfo{year}{2012}{\natexlab{a}}).

\bibitem[{\citenamefont{Chen et~al.}(2012{\natexlab{b}})\citenamefont{Chen,
  Lee, Maity, and Wen}}]{chen-2012b}
\bibinfo{author}{\bibfnamefont{J.-W.} \bibnamefont{Chen}},
  \bibinfo{author}{\bibfnamefont{N.-E.} \bibnamefont{Lee}},
  \bibinfo{author}{\bibfnamefont{D.}~\bibnamefont{Maity}}, \bibnamefont{and}
  \bibinfo{author}{\bibfnamefont{W.-Y.} \bibnamefont{Wen}},
  \bibinfo{journal}{Physics Letters B} \textbf{\bibinfo{volume}{713}},
  \bibinfo{pages}{47} (\bibinfo{year}{2012}{\natexlab{b}}).

\bibitem[{\citenamefont{Barkeshli and Wen}(2010)}]{Barkeshli2010}
\bibinfo{author}{\bibfnamefont{M.}~\bibnamefont{Barkeshli}} \bibnamefont{and}
  \bibinfo{author}{\bibfnamefont{X.-G.} \bibnamefont{Wen}},
  \bibinfo{journal}{Phys. Rev. B} \textbf{\bibinfo{volume}{81}},
  \bibinfo{pages}{155302} (\bibinfo{year}{2010}).

\bibitem[{\citenamefont{Deser et~al.}(1982)\citenamefont{Deser, Jackiw, and
  Templeton}}]{Deser-1982}
\bibinfo{author}{\bibfnamefont{S.}~\bibnamefont{Deser}},
  \bibinfo{author}{\bibfnamefont{R.}~\bibnamefont{Jackiw}}, \bibnamefont{and}
  \bibinfo{author}{\bibfnamefont{S.}~\bibnamefont{Templeton}},
  \bibinfo{journal}{Phys. Rev. Lett.} \textbf{\bibinfo{volume}{48}},
  \bibinfo{pages}{975} (\bibinfo{year}{1982}).

\bibitem[{\citenamefont{Witten}(1988)}]{Witten1988}
\bibinfo{author}{\bibfnamefont{E.}~\bibnamefont{Witten}},
  \bibinfo{journal}{Commun. Math. Phys.} \textbf{\bibinfo{volume}{117}},
  \bibinfo{pages}{353} (\bibinfo{year}{1988}).

\bibitem[{\citenamefont{Witten}(1989)}]{Witten1989}
\bibinfo{author}{\bibfnamefont{E.}~\bibnamefont{Witten}},
  \bibinfo{journal}{Commun. Math. Phys.} \textbf{\bibinfo{volume}{121}},
  \bibinfo{pages}{351} (\bibinfo{year}{1989}).

\bibitem[{\citenamefont{Witten}(2007)}]{witten-2007}
\bibinfo{author}{\bibfnamefont{E.}~\bibnamefont{Witten}},
  \emph{\bibinfo{title}{{Three-Dimensional Gravity Reconsidered}}}
  (\bibinfo{year}{2007}), \eprint{arXiv:0706.3359}.

\bibitem[{\citenamefont{Alvarez-Gaum{\'e}}(1984)}]{Alvarez-Gaume-1984}
\bibinfo{author}{\bibfnamefont{L.}~\bibnamefont{Alvarez-Gaum{\'e}}},
  \bibinfo{journal}{Nucl. Phys. B} \textbf{\bibinfo{volume}{234}},
  \bibinfo{pages}{269} (\bibinfo{year}{1984}).

\bibitem[{\citenamefont{Ryu et~al.}(2012)\citenamefont{Ryu, Moore, and
  Ludwig}}]{Ryu-2012}
\bibinfo{author}{\bibfnamefont{S.}~\bibnamefont{Ryu}},
  \bibinfo{author}{\bibfnamefont{J.~E.} \bibnamefont{Moore}}, \bibnamefont{and}
  \bibinfo{author}{\bibfnamefont{A.~W.~W.} \bibnamefont{Ludwig}},
  \bibinfo{journal}{Phys. Rev. B} \textbf{\bibinfo{volume}{85}},
  \bibinfo{pages}{045104} (\bibinfo{year}{2012}).

\bibitem[{\citenamefont{Stone}(2012)}]{Stone-2012}
\bibinfo{author}{\bibfnamefont{M.}~\bibnamefont{Stone}},
  \bibinfo{journal}{Phys. Rev. B} \textbf{\bibinfo{volume}{85}},
  \bibinfo{pages}{184503} (\bibinfo{year}{2012}).

\bibitem[{\citenamefont{Luttinger}(1964)}]{Luttinger-1964}
\bibinfo{author}{\bibfnamefont{J.~M.} \bibnamefont{Luttinger}},
  \bibinfo{journal}{Phys. Rev.} \textbf{\bibinfo{volume}{135}},
  \bibinfo{pages}{A1505} (\bibinfo{year}{1964}).

\bibitem[{\citenamefont{Hansson et~al.}(1988)\citenamefont{Hansson, Ro\u{c}ek,
  Zahed, and Zhang}}]{Hansson1988Spin}
\bibinfo{author}{\bibfnamefont{T.~H.} \bibnamefont{Hansson}},
  \bibinfo{author}{\bibfnamefont{M.}~\bibnamefont{Ro\u{c}ek}},
  \bibinfo{author}{\bibfnamefont{I.}~\bibnamefont{Zahed}}, \bibnamefont{and}
  \bibinfo{author}{\bibfnamefont{S.~C.} \bibnamefont{Zhang}},
  \bibinfo{journal}{Phys. Lett. B} \textbf{\bibinfo{volume}{214}},
  \bibinfo{pages}{475} (\bibinfo{year}{1988}).

\bibitem[{\citenamefont{Grundberg et~al.}(1989)\citenamefont{Grundberg,
  Hansson, Karlhede, and Lindstr\"{o}m}}]{Grundberg1989}
\bibinfo{author}{\bibfnamefont{J.}~\bibnamefont{Grundberg}},
  \bibinfo{author}{\bibfnamefont{T.~H.} \bibnamefont{Hansson}},
  \bibinfo{author}{\bibfnamefont{A.}~\bibnamefont{Karlhede}}, \bibnamefont{and}
  \bibinfo{author}{\bibfnamefont{U.}~\bibnamefont{Lindstr\"{o}m}},
  \bibinfo{journal}{Phys. Lett. B} \textbf{\bibinfo{volume}{218}},
  \bibinfo{pages}{321} (\bibinfo{year}{1989}).

\bibitem[{\citenamefont{Polyakov}(1988)}]{Polyakov1988}
\bibinfo{author}{\bibfnamefont{A.~M.} \bibnamefont{Polyakov}},
  \bibinfo{journal}{Mod. Phys. Lett. A} \textbf{\bibinfo{volume}{3}},
  \bibinfo{pages}{325} (\bibinfo{year}{1988}).

\bibitem[{\citenamefont{Tze}(1988)}]{Tze1988Manifold}
\bibinfo{author}{\bibfnamefont{C.-H.} \bibnamefont{Tze}},
  \bibinfo{journal}{Int. J. Mod. Phys. A} \textbf{\bibinfo{volume}{3}},
  \bibinfo{pages}{1959} (\bibinfo{year}{1988}).

\bibitem[{\citenamefont{Semenoff and Sodano}(1989)}]{Semenoff1989}
\bibinfo{author}{\bibfnamefont{G.~W.} \bibnamefont{Semenoff}} \bibnamefont{and}
  \bibinfo{author}{\bibfnamefont{P.}~\bibnamefont{Sodano}},
  \bibinfo{journal}{Nucl. Phys. B} \textbf{\bibinfo{volume}{328}},
  \bibinfo{pages}{753} (\bibinfo{year}{1989}).

\bibitem[{\citenamefont{Dunne et~al.}(1989)\citenamefont{Dunne, Jackiw, and
  Trugenberger}}]{Dunne1989}
\bibinfo{author}{\bibfnamefont{G.~V.} \bibnamefont{Dunne}},
  \bibinfo{author}{\bibfnamefont{R.}~\bibnamefont{Jackiw}}, \bibnamefont{and}
  \bibinfo{author}{\bibfnamefont{C.~A.} \bibnamefont{Trugenberger}},
  \bibinfo{journal}{Ann. Phys.} \textbf{\bibinfo{volume}{194}},
  \bibinfo{pages}{197} (\bibinfo{year}{1989}).

\bibitem[{\citenamefont{You et~al.}(2014)\citenamefont{You, Cho, and
  Fradkin}}]{You_unpublished}
\bibinfo{author}{\bibfnamefont{Y.}~\bibnamefont{You}},
  \bibinfo{author}{\bibfnamefont{G.~Y.} \bibnamefont{Cho}}, \bibnamefont{and}
  \bibinfo{author}{\bibfnamefont{E.}~\bibnamefont{Fradkin}}
  (\bibinfo{year}{2014}), \bibinfo{note}{in preparation}.

\bibitem[{\citenamefont{Weinberg}(2005)}]{weinberg}
\bibinfo{author}{\bibfnamefont{S.}~\bibnamefont{Weinberg}},
  \emph{\bibinfo{title}{{The Quantum Theory of Fields}}}
  (\bibinfo{publisher}{Cambridge University Press},
  \bibinfo{address}{Cambridge, UK}, \bibinfo{year}{2005}).

\bibitem[{\citenamefont{Elitzur et~al.}(1989)\citenamefont{Elitzur, Moore,
  Schwimmer, and Seiberg}}]{Elitzur1989}
\bibinfo{author}{\bibfnamefont{S.}~\bibnamefont{Elitzur}},
  \bibinfo{author}{\bibfnamefont{G.}~\bibnamefont{Moore}},
  \bibinfo{author}{\bibfnamefont{A.}~\bibnamefont{Schwimmer}},
  \bibnamefont{and} \bibinfo{author}{\bibfnamefont{N.}~\bibnamefont{Seiberg}},
  \bibinfo{journal}{Nuclear Physics B} \textbf{\bibinfo{volume}{326}},
  \bibinfo{pages}{108} (\bibinfo{year}{1989}).

\bibitem[{\citenamefont{Barkeshli and McGreevy}(2014)}]{Barkeshli2012}
\bibinfo{author}{\bibfnamefont{M.}~\bibnamefont{Barkeshli}} \bibnamefont{and}
  \bibinfo{author}{\bibfnamefont{J.}~\bibnamefont{McGreevy}},
  \bibinfo{journal}{Phys. Rev. B} \textbf{\bibinfo{volume}{89}},
  \bibinfo{pages}{235116} (\bibinfo{year}{2014}).

\end{thebibliography}

\end{document}